\documentclass[twocolumn,prd,amsmath,amssymb, aps,superscriptaddress, nofootinbib]{revtex4-1}
\usepackage[version=4]{mhchem} 
\usepackage[normalem]{ulem}
\usepackage{amsmath}
\usepackage{amssymb}
\usepackage{amsfonts}
\usepackage{mathrsfs}
\usepackage{graphicx}
\usepackage{xcolor}
\usepackage{xfrac}
\usepackage{footmisc}
\usepackage{pifont}
\usepackage{times}
\usepackage{epstopdf}
\usepackage{enumerate}
\usepackage[hidelinks]{hyperref}
\usepackage{enumitem}
\usepackage{multirow}
\usepackage{booktabs}
\usepackage{slashed}
\usepackage{relsize}
\usepackage{epsfig}
\usepackage{verbatim}
\usepackage{bm}
\usepackage[utf8]{inputenc}
\usepackage{graphics}
\usepackage{subfigure}
\usepackage[english]{babel}
\usepackage{orcidlink}

\definecolor{zima_blue}{HTML}{1393C1}
\hypersetup{setpagesize=false,
bookmarksnumbered=true,
bookmarksopen=true, 
colorlinks=true,
linkcolor=zima_blue,
urlcolor=zima_blue,
citecolor=zima_blue,
linktocpage=false}

\usepackage{array}
\newcolumntype{P}[1]{>{\centering\arraybackslash}p{#1}}
\newcolumntype{M}[1]{>{\centering\arraybackslash}m{#1}}
\newcommand{\VEV}[1]{\left\langle #1 \right\rangle}

\begin{document}

\title{Resonant ALP-Portal Dark Matter Annihilation as a Solution to the $B^{\pm} \to K^{\pm} \nu \bar{\nu}$ Excess
}

\author{Kewen Ding}
\email{dlsc52000@s.ytu.edu.cn}
\affiliation{Department of Physics, Yantai University, Yantai 264005, China}

\author{Ying Li} 
\email{liying@ytu.edu.cn}
\affiliation{Department of Physics, Yantai University, Yantai 264005, China}

\author{Xuewen Liu~\orcidlink{0000-0003-3652-7237}}
\email{xuewenliu@ytu.edu.cn}
\affiliation{Department of Physics, Yantai University, Yantai 264005, China}

\author{Yu Liu} 
\email{liuyu2020@s.ytu.edu.cn}
\affiliation{School of Physics, Henan Normal University, Xinxiang 453007, China}

\author{Chih-Ting Lu} 
\email{ctlu@njnu.edu.cn}
\affiliation{Department of Physics and Institute of Theoretical Physics, Nanjing Normal University, Nanjing, 210023, China}
\affiliation{Nanjing Key Laboratory of Particle Physics and Astrophysics, Nanjing, 210023, China}

\author{Bin Zhu} 
\email{zhubin@mail.nankai.edu.cn}
\affiliation{Department of Physics, Yantai University, Yantai 264005, China}

\begin{abstract}
 
The Belle II collaboration recently reported a $2.7\sigma$ excess in the rare decay $B^\pm \to K^\pm \nu \bar{\nu}$, potentially signaling new physics. We propose an axion-like particle (ALP)-portal dark matter (DM) framework to explain this anomaly while satisfying the observed DM relic abundance. By invoking a resonant annihilation mechanism ($m_a \sim 2m_\chi$), we demonstrate that the ALP-mediated interactions between the Standard Model and DM sectors simultaneously account for the $B^\pm \to K^\pm \nu \bar{\nu}$ anomaly and thermal freeze-out dynamics. Two distinct scenarios-long-lived ALPs decaying outside detectors (displaced diphotons) and ALPs decaying invisibly to DM pairs (missing energy)-are examined. While the displaced diphotons scenario is excluded by kaon decay bounds ($K^\pm \to \pi^\pm + \text{inv.}$), the invisible decay channel remains unconstrained and aligns with Belle II's missing energy signature. 
Using the coupled Boltzmann equation formalism, we rigorously incorporate early kinetic decoupling effects, revealing deviations up to a factor of 20 from traditional relic density predictions in resonance regions. 
For the missing energy scenario, the viable parameter space features ALP-SM and ALP-DM couplings: $g_{aWW}(g_{a\gamma\gamma}) \in (7.13 \times 10^{-5} - 9.60 \times 10^{-5})\, \text{GeV}^{-1}$ (from $B^\pm \to K^\pm a$) and $g_{a\chi\chi} \in (7.12\times10^{-5} - 7.73\times 10^{-3})\, \text{GeV}^{-1}$ (for resonant annihilation), accommodating ALP masses $m_a \in (0.6, 4.8)\, \text{GeV}$. Therefore, this work establishes the ALP portal as a viable bridge between the $B^\pm \to K^\pm \nu \bar{\nu}$ anomaly and thermal DM production, emphasizing precision calculations of thermal decoupling in resonance regimes. 

\end{abstract}

\maketitle

\section{Introduction}
\label{sec-intro}

Axion-like particles (ALPs) can be naturally generated through the breaking of a global $U(1)$ symmetry~\cite{Graham:2015ouw,Choi:2020rgn}. ALPs were originally proposed to solve the strong CP problem~\cite{Peccei:1977hh,Weinberg:1977ma,Wilczek:1977pj,Kim:1979if}. Later, they became compelling candidates for dark matter (DM) via the misalignment mechanism~\cite{Preskill:1982cy,Abbott:1982af,Dine:1982ah}. Additionally, ALPs could influence the dynamics of inflation and reheating~\cite{Blumenhagen:2014gta,Kobayashi:2020ryx}, as well as the generation of baryon asymmetry through CP-violating processes in the early Universe~\cite{Guendelman:1991se,Domcke:2020kcp,Co:2024oek}. Moreover, ALPs can act as mediators between the visible and dark sectors, providing a portal to explore the dark sector, often referred to as the ALP portal~\cite{Nomura:2008ru}. Given their potential to address major unresolved problems and bridge various domains of physics from particle physics to cosmology and astrophysics, ALPs have become a highly attractive focus of modern theoretical and experimental research.

In this work, we focus on ALP-portal DM models. Since the couplings of ALPs to a DM pair and to pairs of Standard Model (SM) particles are naturally suppressed by a cutoff scale associated with global $U(1)$ symmetry breaking, ALP-portal DM models allow for diverse production mechanisms, including the freeze-out~\cite{Bharucha:2022lty,Allen:2024ndv}, SIMP mechanism \cite{Hochberg:2018rjs} and freeze-in~\cite{Im:2019iwd,Ghosh:2023tyz,Arias:2025nub} mechanisms. These mechanisms enable ALP-portal DM models to accommodate a broad range of coupling strengths and DM masses. Notably, when the ALP mass aligns with the DM annihilation resonance, it significantly enhances the annihilation cross-section~\cite{Guo:2009aj,Laine:2022ner}. When a pair of DM particles predominantly annihilates to photons, this feature provides a natural explanation for specific astrophysical signals, such as excesses observed in cosmic-ray or gamma-ray data, and allows precise tuning of DM properties~\cite{Lee:2012bq,Ibarra:2013eda,Klangburam:2023vjv,Allen:2024ndv,Yang:2024jtp}. Furthermore, ALP-portal DM models exhibit rich experimental phenomenology, including searches in collider~\cite{Biswas:2019lcp,Armando:2023zwz} and beam-dump~\cite{Ghosh:2023tyz,Armando:2023zwz,Jodlowski:2024lab} experiments.

On the other hand, recently, Belle II reported the first evidence of the \(B^\pm \rightarrow K^\pm \nu\bar{\nu}\) decay channel with a significance of \(3.5\sigma\)~\cite{Belle-II:2023esi}. However, a \(2.7\sigma\) deviation from the SM prediction has also been announced in this process. This deviation may indicate the presence of new physics~\cite{Athron:2023hmz,Bause:2023mfe,Allwicher:2023xba,Abdughani:2023dlr,He:2023bnk,Datta:2023iln,Altmannshofer:2023hkn,McKeen:2023uzo,Wang:2023trd,Fridell:2023ssf,Ho:2024cwk,Chen:2024jlj,Gabrielli:2024wys,Hou:2024vyw,Chen:2024cll,He:2024iju,Bolton:2024egx,Marzocca:2024hua,Rosauro-Alcaraz:2024mvx,Kim:2024tsm,Hati:2024ppg,Kolay:2024wns,Allwicher:2024ncl,Becirevic:2024iyi,Altmannshofer:2024kxb,Buras:2024mnq,Hu:2024mgf,Zhang:2024hkn,Calibbi:2025rpx,Lee:2025jky,He:2025jfc,Berezhnoy:2025tiw,Bolton:2025fsq,Aliev:2025hyp,Chen:2025npb,Kolay:2025jip}, and ALPs, as candidates for physics beyond the SM, could potentially explain the excess reported by Belle II~\cite{Altmannshofer:2024kxb,Calibbi:2025rpx}. By further investigating the contributions of ALPs to this decay channel, we can evaluate whether they are responsible for the observed excess.

In the thermal freeze-out framework, several DM annihilation scenarios exist in the ALP-portal model, depending on the mass spectrum between the DM and the ALP, as well as the size of coupling constants. In this study, we aim to explore the potential connection between the explanation of the $B^{\pm} \to K^{\pm} \nu \bar{\nu}$ excess and the DM relic abundance within the ALP-portal DM model. For the DM $p$-wave or forbidden annihilation scenario, the ALP predominantly decays into diphotons, and ALPs are required to be long-lived particles (LLPs) that decay outside the detectors, with the signature resulting in missing energy, which can explain this excess. In the DM $s$-wave annihilation scenario, despite the strict constraints imposed by the Cosmic Microwave Background and indirect detection experiments, these limitations can be effectively circumvented if DM annihilation during freeze-out is resonantly enhanced~\cite{Wang:2025tdx,Belanger:2025kce}. In this case, the ALP can predominantly decay into a pair of DM particles, which directly accounts for this excess. Therefore, the $B^{\pm} \to K^{\pm} \nu \bar{\nu}$ excess observed at Belle II could provide a hint for the possible DM annihilation scenario in the early Universe.

Among these DM annihilation scenarios, special care must be taken in calculating the resonant $s$-wave annihilation. In this work, we employ the coupled Boltzmann equation (CBE) method, accurately accounting for the resonance mechanism, in which the early kinetic decoupling (EKD) effect has been taken into account. The EKD effect has been extensively studied in the literature~\cite{vandenAarssen:2012ag,Binder:2017rgn,Brummer:2019inq,Ala-Mattinen:2019mpa,Abe:2020obo,Binder:2021bmg,Zhu:2021vlz,Hryczuk:2021qtz,Abe:2021jcz,Du:2021jcj,Ala-Mattinen:2022nuj,Hryczuk:2022gay,Liu:2023kat,Aboubrahim:2023yag,Duan:2024urq} to calculate the DM relic density. 
In the context of the Boltzmann equation, it is generally imperative to consider the scattering collision terms for the resonant mechanism. The momentum transfer rate associated with elastic scattering processes, which are crucial for maintaining thermal equilibrium, is significantly suppressed relative to the annihilation rate. This suppression arises due to the small coupling constant required to meet the relic density criterion, absent any enhancement of the scattering rate. Consequently, kinetic equilibrium decouples prematurely and the outcomes derived from the CBE method diverge significantly from those obtained through the standard calculation formulated by Gondolo and Gelmini \cite{Gondolo:1990dk} (denoted as the `NBE' method since this treatment involves solving the Boltzmann equations of number density)). Our investigation reveals that the observed relic abundance of DM can be produced within a precisely defined parameter space region. This region represents a promising avenue for further exploration at the Belle II experiment, particularly in the context of explaining the excess observed in the $B^{\pm} \to K^{\pm} \nu \bar{\nu}$ decay mode.

The structure of the rest of this paper is organized as follows: In Sec.~\ref{sec:ALPs}, we first review the ALP-portal DM model and its thermal annihilation mechanisms. In Sec.~\ref{sec:BelleII}, we discuss exotic $B$ meson decays that produce ALPs and explore their connection to the $B^{\pm} \to K^{\pm} \nu \bar{\nu}$ anomaly at Belle II. In Sec.~\ref{sec:resonance}, we present precise calculations of the DM relic density in the resonance annihilation scenario. In Sec.~\ref{sec:results}, we display our combined results for the parameter space that explains the $B^{\pm} \to K^{\pm} \nu \bar{\nu}$ excess and yields the correct DM relic density. Finally, we summarize our findings in Sec.~\ref{sec:conclusion}.

\section{ALP portal DM model and its annihilation mechanisms}
\label{sec:ALPs}

In this section, we begin by reviewing the ALP portal DM model and the effective interactions of ALPs. Next, we discuss the decay patterns of ALPs. Finally, we explore various thermal DM production mechanisms within this model.

\subsection{ALP portal DM model}
For the interactions of a pseudoscalar ALP, $a$, with SM gauge bosons, we consider the Lagrangian
\begin{eqnarray}
\mathcal{L}_{\text{EW}}&=& \frac{1}{2} \partial^\mu a \, \partial_\mu a - \frac{1}{2}m^{2}_{a} \, a^2 - \frac{c_B}{4 \, f_a} \, a \, B^{\mu\nu}\tilde{B}_{\mu\nu} \nonumber\\
&-& \frac{c_W}{4 \, f_a} \, a \, W^{i,\mu\nu}\tilde{W}^i_{\mu\nu} \; ,
\label{eq:Lag}
\end{eqnarray}
where $B^{\mu\nu}$ and $W^{i,\mu\nu}$ denote the field strength tensors of $U(1)_Y$ and $SU(2)_L$ gauge groups before electroweak symmetry breaking (EWSB), respectively, and the dual field strength tensors are defined via $\tilde{B}_{\mu\nu} = \frac{1}{2} \epsilon_{\mu\nu\rho\sigma} \, B^{\rho\sigma}$. The parameters $c_B,c_W$ are the corresponding Wilson coefficients for ALP-gauge boson interactions. 
$m_a$ and $f_a$ denote the ALP mass and decay constant, which are taken to be independent parameters.

After EWSB, the last two terms in Eq.~(\ref{eq:Lag}) induce four different interactions between the ALP and SM gauge bosons:
\begin{eqnarray}
     \mathcal{L}_{\text{EW}} &\supset& - \frac{g_{a\gamma\gamma}}{4} a F_{\mu\nu} \tilde{F}^{\mu\nu} - \frac{g_{a\gamma Z}}{4} a F_{\mu\nu} \tilde{Z}^{\mu\nu} \nonumber\\ 
     &-& \frac{g_{aZZ}}{4} a Z_{\mu\nu} \tilde{Z}^{\mu\nu}
     -\frac{g_{aWW}}{4} a W_{\mu\nu} \tilde{W}^{\mu\nu} \; ,
\end{eqnarray}
where $F_{\mu\nu}$, $Z_{\mu\nu}$, and $W_{\mu\nu}$ are the field strength tensors of the photon, $Z$, and $W^{\pm}$ bosons, respectively, and their duals have been defined as above. As pointed out in Ref.~\cite{Izaguirre:2016dfi}, the coupling of ALPs to $W^\pm$ bosons can induce flavor-changing processes such as $B \to K a$ and offer the best sensitivity to ALPs with masses below $5$ GeV. Therefore, the two most relevant couplings in this study are those of the ALP to a pair of photons and a pair of $W$ bosons, which are expressed in terms of $c_B$ and $c_W$~\cite{Georgi:1986df,Brivio:2017ije,Ren:2021prq,Lu:2023ryd,Cheung:2023nzg}, 
\begin{equation}
 g_{a\gamma\gamma} = \frac{c_B \, \cos^2 \theta_\mathrm{W} + c_W \, \sin^2 \theta_\mathrm{W}}{f_a} \; , 
\end{equation}
\begin{equation}
g_{aWW} = \frac{c_W }{f_a}  \; , 
\end{equation}
where $\theta_\mathrm{W}$ denotes the Weinberg angle. In this study, we choose $c_B=c_W$ as a benchmark model.

\begin{figure}[h]
  \centering
  \includegraphics[width=0.47\textwidth]{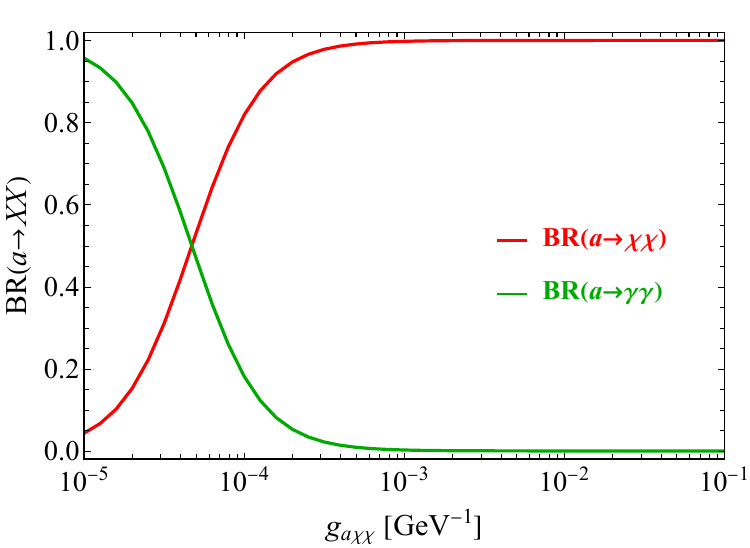}  
  \caption{Branching ratios for two ALP decay channels, $a \to \gamma\gamma$ and $a \to \chi\bar{\chi}$. Here we take $m_a=1$ GeV, $m_\chi=0.475 m_a$, $g_{a\gamma\gamma}= 10^{-4}$ GeV$^{-1}$ as a benchmark point and vary the value of $g_{a\chi\chi}$. }
  \label{fig:BR}
\end{figure}

In addition to interactions between the ALP and gauge bosons, we also consider the ALP interacting with the dark sector via the ALP portal~\cite{Dolan:2017osp,Gola:2021abm,Bharucha:2022lty,Ghosh:2023tyz}, 
\begin{align}
    \mathcal{L}_{\text{DM}} = \bar{\chi}i\partial_{\mu}\gamma^{\mu}\chi -m_{\chi}\bar{\chi^c}\chi
 -g_{a\chi\chi}\partial_{\mu}a\Bar{\chi}\gamma^{\mu}\gamma^5\chi 
\end{align}
where $\chi$ represents the Majorana fermionic dark matter, and $g_{a\chi\chi}$ is the coupling of the ALP to a pair of DM particles, with dimensions of inverse mass.

For an $\mathcal{O}(1)$ GeV ALP, it predominantly decays into either a pair of photons or a pair of DM particles, depending on the mass relation between the ALP and DM. The partial decay width for the process $a \to \gamma\gamma$ is given by:   
\begin{equation}
\Gamma_{a\gamma} = \frac{g_{a\gamma\gamma}^2 \, m_{a}^3}{64 \pi} \;.
\end{equation} 
Additionally, if $m_a > 2m_\chi$, the partial decay width for $a \to \chi\bar{\chi}$ must also be included: 
\begin{equation}
\Gamma_{a\chi} = \frac{g_{a\chi\chi}^2 \, m_{a} m_\chi^2}{\pi} \sqrt{1-\frac{4m_\chi^2}{m_a^2}} \;.
\end{equation}
To illustrate the interplay between the two ALP decay channels, $a \to \gamma\gamma$ and $a \to \chi\bar{\chi}$, for varying $g_{a\chi\chi}$, we fix $m_a = 1$ GeV, $m_\chi = 0.475 m_a$, and $g_{a\gamma\gamma} = 10^{-4}$ GeV$^{-1}$ as a benchmark point, and present the results in Fig.~\ref{fig:BR}. It is evident that when $g_{a\chi\chi} \gtrsim 10^{-4}$ GeV$^{-1}$, the decay channel $a \to \chi\bar{\chi}$ dominates. Otherwise, $a \to \gamma\gamma$ becomes the primary decay mode. 

\subsection{Thermal DM annihilation mechanisms }

DM annihilations to photons in this model can be dominant by the $s$-wave contribution via $a$-resonance ($\chi\overline{\chi}\to a\to\gamma\gamma$), or $p$-wave contribution ($\chi\overline{\chi}\to a a$). 
In addition, some different mass relations between $\chi$ and $a$ can alter the velocity dependence of the DM annihilation cross-section.  
For example, in the $\chi\overline{\chi}\to a a$ cases, the main processes involve $p$-wave annihilation (velocity-squared suppression) and forbidden annihilation (velocity-exponential suppression). On the other hand, when $2 m_\chi\approx m_a$, the decay channel $a\to \chi\bar{\chi} $ opens and the resonant $s$-wave annihilation becomes dominant.

The possible annihilation mechanisms based on the relationships between $m_\chi$ and $m_a$ appear as what follows: 
\begin{enumerate}
    \item $p$-wave annihilation ($m_a < m_{\chi}$): The dominant DM annihilation channel is $\chi\overline{\chi}\to aa$ which is $p$-wave and then each ALP decays into diphoton.
    \item forbidden annihilation ($m_a\gtrsim m_{\chi}$): The annihilation process is also $\chi\overline{\chi}\to aa$, albeit with $m_\chi\lesssim m_a$. Essentially, forbidden annihilations cease when the kinetic energies of DM particles fall below threshold energies. Here each ALP also decays into diphoton.
    \item $s$-wave annihilation ($m_a > 2m_{\chi}$ or $2m_{\chi} > m_a > m_{\chi}$): The annihilation process is $\chi\overline{\chi}\to a\to\gamma\gamma$, and both $g_{a\gamma\gamma}$ and $g_{a\chi\chi}$ must be sufficiently large to account for the DM relic abundance~\cite{Bharucha:2022lty,Allen:2024ndv}. Additionally, in this scenario, the ALP predominantly decays into either diphoton or a pair of DM particles, which can be probed in various accelerator~\cite{Biswas:2019lcp,Armando:2023zwz,Ghosh:2023tyz,Jodlowski:2024lab} and indirect detection~\cite{Lee:2012bq,Ibarra:2013eda,Klangburam:2023vjv,Allen:2024ndv,Yang:2024jtp} experiments. Consequently, the parameter space for this scenario is already subject to significant constraints. 
    \item Resonance annihilation ($m_a \sim 2 m_{\chi} $): The annihilation cross-section for $\chi\overline{\chi}\to\gamma\gamma$ can experience significant enhancement through an $a$-resonance. We focus on the scenario where DM predominantly couples to photons. Strong constraints on this scenario, imposed by the CMB and indirect detection experiments can be circumvented if DM annihilations during freeze-out are resonantly enhanced~\cite{Wang:2025tdx,Belanger:2025kce}. In this context, we find that the observed DM relic abundance can be reproduced within a well-defined region of parameter space that is currently consistent with all experimental constraints. However, this parameter space can be explored in the coming years through invisible ALPs searches at the Belle II~\cite{Dolan:2017osp,Acanfora:2023gzr}. To ensure accuracy in our analysis, we employ the coupled Boltzmann equation method~\cite{Liu:2023kat,Zhu:2021vlz} to calculate the DM relic density, properly accounting for the resonance mechanism.
\end{enumerate}

\section{ALPs from exotic $B$ meson decays and the $B^{\pm}\to K^{\pm}\nu\bar{\nu}$ anomaly at Belle II}
\label{sec:BelleII}

The ALPs are expected to be produced via flavor-changing processes at $B$-factories, specifically focusing on their generation during $B$ meson decays. We examine the two-body decay process $B^\pm\rightarrow K^\pm a$.  At energy scales below the weak interaction scale, the relevant effective Lagrangian is~\cite{Izaguirre:2016dfi,Ferber:2022rsf}
\begin{eqnarray}
    \mathcal{L}_{\text{eff}} &\supset& -g_{aWW}  C_{sb}  \partial^\mu a (\bar{s}_L \gamma_\mu b_L) + \text{h.c.} .
\end{eqnarray}
The Wilson coefficient $C_{sb}$  characterizes the ALPs-mediated $b \to s$ flavor-changing process. 
The partial decay width for this process can be written as
\begin{eqnarray}
\Gamma(B^\pm \rightarrow K^\pm a)&=&\frac{g^2_{aWW}}{64\pi m^3_{B^\pm}}|C_{sb}|^2 F^2_K(m_a)(m^2_{B^\pm}-m^2_{K^\pm})^2\nonumber\\
&\times&\lambda^{1/2}(m^2_{B^\pm},m^2_{K^\pm},m^2_a),
\end{eqnarray}
where $m_{B^\pm}$, $m_{K^\pm}$ represent the masses of the $B^\pm$, $K^\pm$ mesons, respectively. The scalar form factor $F_K(q)$ for $K^\pm$ characterizes the influence of momentum transfer $q^2$ on the decay process, which is taken from \cite{Gubernari:2023puw} with associated uncertainties. The Wilson coefficient $C_{sb}$ is expressed as
\footnote{Note that the Wilson coefficient expression matches Eq.~(2.4) in~\cite{Ferber:2022rsf} when $c_{tt}(\Lambda) = 0$, corresponding to the absence of direct ALP-fermion couplings in our Lagrangian. Furthermore, gauge-boson-loop-induced ALP-fermion couplings (estimated in Eq.~(2.15) of~\cite{Bauer:2021mvw}) are suppressed by at least three orders of magnitude relative to direct couplings. Consequently, these induced couplings and associated running effects remain phenomenologically negligible and do not impact our conclusions.}
\begin{equation}
    C_{sb}=-\frac{3 m^2_W}{16 \pi^2 v^2} V^*_{ts}V_{tb}f\left(\frac{m^2_t}{m^2_W}\right),
\end{equation}
where $m_W$ and $m_t$ are masses of the $W$-boson and the top quark, $v$ is the vacuum expectation value of the Higgs field, and $V^*_{ts}$ and $V_{tb}$ are elements of the CKM matrix.

Additionally, we introduce two key mathematical functions to represent the decay dynamics and their integrals. These include kinematic function: 
\begin{equation}
\lambda(\alpha ,\beta,\gamma)=\alpha^2+\beta^2+\gamma^2-2(\alpha\beta+\alpha\gamma+\beta\gamma),  
\end{equation} 
and the integral function: 
\begin{equation}
f(x)=x[1+x(\log x-1)]/(1-x)^2, 
\end{equation} 
which facilitate the analysis of the decay mechanisms.

The detection of ALPs in this process exhibits three distinct signatures, depending on the ALP decay length in the laboratory frame: (1) prompt diphoton, (2) displaced diphoton, (3) missing energy. The BaBar collaboration has already utilized the first two signatures, prompt and displaced diphotons, to search for ALPs, particularly in the mass range $0.175~{\rm GeV} < m_a < 4.78~{\rm GeV}$, where the coupling constant $g_{aWW}\sim 10^{-5}$ GeV$^{-1}$~\cite{BaBar:2021ich}. However, due to significant background contamination from $\eta$ and $\eta^\prime$ mesons, the mass ranges $0.45 ~{\rm GeV} < m_a < 0.63~{\rm GeV}$ and $0.91~{\rm GeV} < m_a < 1.01~{\rm GeV}$ are excluded from this search. These regions are challenging because the $\eta$ and $\eta^\prime$ mesons contribute substantially to the background, making it difficult to distinguish the ALP signal.

To overcome these limitations, a novel search method using photon-jets as a distinctive signature for ALPs has been proposed at the Large Hadron Collider (LHC). As detailed in Refs.~\cite{Wang:2021uyb,Ren:2021prq}, this approach exploits the decay of ALPs into pairs of highly collimated photons, forming a photon-jet, a signature that becomes prominent at LHC energies. Crucially, photon-jets differ fundamentally from QCD jets: nearly all of their energy is deposited in the electromagnetic calorimeter (ECAL), whereas QCD jets deposit most of their energy in the hadronic calorimeter (HCAL). Since light ALPs decay predominantly into photon-jets detectable in the ECAL, this strategy remains effective even when the ALP mass approaches that of the $\eta$ and $\eta^{\prime}$ mesons, which are primarily produced via QCD processes. This feature allows for significant mitigation of $\eta$ and $\eta^{\prime}$ induced background contamination, providing a powerful experimental approach for ALP searches. Combining LHC photon-jet analyses with existing $B$-factory studies can further expand the ALP discovery potential and improve constraints on their mass and coupling constants, thereby enhancing the probe of new physics beyond the SM.

We also identify the mass range $0.1~{\rm GeV} < m_a < 0.175~{\rm GeV}$ as a blind spot for ALP prompt decay to diphoton searches due to significant background contamination from $\pi^0$ meson decays. This contamination complicates the separation of ALP signals from background noise, making it challenging to detect ALPs in this mass range using current experimental techniques. However, for the long-lived ALP in this blind spot, the ALPs decay length would be sufficiently large that it does not produce any visible decay products within the detector of $B$ factories. As a result, the ALP would manifest as a missing energy signature. Therefore, we can employ missing energy searches to constrain the potential contribution of ALPs. 
Notably, this method can be applied to other ALP mass range when light ALPs become very long-lived. Specifically, we propose leveraging the existing $B^\pm \rightarrow K^\pm \nu\bar{\nu}$ 
process to limit the presence of ALPs indirectly. Since the neutrino pair is electrically neutral and difficult to detect, this decay process appears as missing energy in the detectors. If the ALP decays outside the detector, its signature would resemble the missing energy feature of the $B^\pm \rightarrow K^\pm \nu\bar{\nu}$ process. By accurately measuring this process, we can place upper limits on the ALPs contribution.

Additionally, we consider a complementary scenario in which the ALP primarily decays into a pair of DM particles which makes the ALP decay products invisible to the detectors. As discussed above, distinguishing this scenario from the $B^\pm \rightarrow K^\pm \nu\bar{\nu}$ process in the SM is challenging. Therefore, precision measurements of the $B^\pm \rightarrow K^\pm \nu\bar{\nu}$ process can also place upper limits on the ALP contribution in this scenario.

Notably, the Belle II experiment has searched for the $B^\pm \rightarrow K^\pm \nu\bar{\nu}$ decay and observed a $2.7\sigma$ deviation from the SM prediction of BR$(B^{\pm}\to K^{\pm}\nu\bar{\nu})=(5.58\pm 0.37)\times 10^{-6}$~\cite{Parrott:2022zte}. In contrast, the measured branching ratio is BR$(B^{\pm}\to K^{\pm}\nu\bar{\nu})=[2.3\pm 0.5(\text{stat})^{+0.5}_{-0.4}(\text{syst})]\times 10^{-5}$~\cite{Belle-II:2023esi}. Although it is premature to assert whether this excess results from statistical fluctuations or is a genuine indication of new physics, it does represent the first evidence of the $B^\pm \rightarrow K^\pm \nu\bar{\nu}$ process with a significance of $3.5\sigma$. Therefore, further accumulation of events and improvements in measurements are urgently needed. It is also valuable to consider possible new physics explanations for this observed anomaly while awaiting the next update on the $B^\pm \rightarrow K^\pm \nu\bar{\nu}$ process. In fact, several proposals have already been put forward to explain this excess, including new $U(1)$ extensions of the SM~\cite{Athron:2023hmz,Buras:2024mnq}, the SMEFT framework~\cite{Allwicher:2023xba,Chen:2024jlj,Hou:2024vyw,Marzocca:2024hua,Rosauro-Alcaraz:2024mvx}, light dark matter models~\cite{Abdughani:2023dlr,He:2023bnk,McKeen:2023uzo,Fridell:2023ssf,Ho:2024cwk,He:2024iju,He:2025jfc,Berezhnoy:2025tiw}, ALPs~\cite{Altmannshofer:2024kxb,Calibbi:2025rpx}, and other possibilities~\cite{Datta:2023iln,Altmannshofer:2023hkn,Gabrielli:2024wys,Chen:2024cll,Hati:2024ppg,Becirevic:2024iyi,Hu:2024mgf,Zhang:2024hkn,Bolton:2025fsq,Aliev:2025hyp,Chen:2025npb}.

In this study, we aim to use the ALP-portal DM model to explain the observed $B^\pm \rightarrow K^\pm \nu\bar{\nu}$ excess from the Belle II experiment. As discussed in Sec.~\ref{sec:ALPs}, ALPs predominantly decay into diphotons in both the DM $p$-wave annihilation and forbidden annihilation scenarios. In these scenarios, ALPs are required to be light, and the coupling $g_{a\gamma\gamma}$ must be small enough to ensure that the ALP lifetime is long enough for them to travel outside the detectors, with the final state resulting in missing energy. Alternatively, in the DM $s$-wave annihilation scenario, ALPs can decay into diphotons and a pair of DM particles, as shown in Fig.~\ref{fig:BR}, depending on the relative magnitudes of $g_{a\gamma\gamma}$ and $g_{a\chi\chi}$. When the invisible decay channel of ALPs dominates, the lifetime of ALPs becomes less of a concern, as the final state already results in missing energy. By further investigating the contribution of ALPs to the $B^\pm \rightarrow K^\pm \nu\bar{\nu}$ excess in Sec.~\ref{sec:results}, we can assess whether they could be the source of this observed excess.

It should be noted that this analysis provides an approximate approach to illustrate the potential compatibility of the ALP-mediated process with the excess. The branching ratio extracted by Belle II relies on SM kinematics, while the signal morphology in our model could introduce non-trivial effects in the $q^2$ spectrum. As indicated in Refs.~\cite{Altmannshofer:2023hkn,Bolton:2024egx,Bolton:2025fsq,Abumusabh:2025zsr}, these effects can serve as a probe for the decay $B^+ \to K^+ a$. A definitive assessment of our scenario's compatibility with the Belle II data would require a full re-fit of the experimental distributions. Such a dedicated study, however, is beyond the scope of this work and is postponed for future research.

\section{Precision DM relic density in the resonance mechanism}
\label{sec:resonance}

As extensively discussed in the literature, the freeze-out dynamics of DM exhibit significant complexity within the resonance region of annihilation~\cite{vandenAarssen:2012ag,Binder:2017rgn,Brummer:2019inq,Ala-Mattinen:2019mpa,Abe:2020obo,Binder:2021bmg,Zhu:2021vlz,Hryczuk:2021qtz,Abe:2021jcz,Du:2021jcj,Ala-Mattinen:2022nuj,Hryczuk:2022gay,Liu:2023kat,Aboubrahim:2023yag,Duan:2024urq}. In this section, we provide a comprehensive review of the coupled Boltzmann equations, which play a crucial role in accurately calculating the relic density. Additionally, we explore in detail the early kinetic decoupling (EKD) effect that emerges in the  resonant annihilation.

\subsection{Coupled Boltzmann Equation}
Keeping kinetic equilibrium during and even after the freeze-out period is a key assumption for standard relic density calculations. Nevertheless, this assumption is not invariably valid, especially in contexts where kinetic decoupling occurs prior to chemical decoupling.
To investigate the DM relic density considering the impact of early kinetic decoupling, 
we should focus on the following Boltzmann equation for DM phase-space distribution~\cite{Binder:2017rgn,Binder:2021bmg}:  
\begin{align}
 E \left( \frac{\partial}{\partial t} - H \vec{p}\cdot \frac{\partial}{\partial \vec{p}} \right) f_\chi (t, \vec{p})=
C_{\rm ann.}[f_\chi] + C_{\rm el.}[f_\chi],
\label{Eq:boltzmann}
\end{align}
where $E$ is the energy of the DM, $H$ is the Hubble constant, $\vec{p}$ is the momentum of DM,
and $f_\chi$ is the DM phase-space density.
The collision terms, $C_{\rm ann.}$ and $C_{\rm el.}$, account for the annihilation of DM particles into particles of the thermal bath and the elastic scattering processes between DM and SM particles, respectively.
For two-body processes, 

\begin{eqnarray}
 C_{\rm ann.}&=&\frac{1}{2 g_\chi}
\sum
\int \frac{d^3 p'}{(2\pi)^3 2 E_{p'}}
\int \frac{d^3 k}{(2\pi)^3 2 E_k}
\int \frac{d^3 k'}{(2\pi)^3 2 E_{k'}}\nonumber\\
&\times&  
(2\pi)^4 \delta^4(p + p' - k - k') \nonumber\\
&\times& 
\Bigl(-|{\cal M}_{\chi \chi \to {\cal B} {\cal B}'}|^2 f_\chi(\vec{p}) f_\chi(\vec{p'}) (1 \pm f^{\rm eq}_{\cal B}(\vec{k}))\nonumber\\
&\times&(1 \pm f^{\rm eq}_{\cal B'}(\vec{k'}))
+|{\cal M}_{{\cal B}{\cal B'} \to \chi \chi}|^2 f^{\rm eq}_{\cal B}(\vec{k}) f^{\rm eq}_{{\cal B}'}(\vec{k'}) \nonumber\\
&\times&(1 \pm f_\chi(\vec{p})) (1 \pm f_\chi(\vec{p'}))
\Bigr),
\end{eqnarray}
\begin{eqnarray}
 C_{\rm el.}
&=&
\frac{1}{2g_\chi}
\sum
\int \frac{d^3 p'}{(2\pi)^3 2 E_{p'}}
\int \frac{d^3 k}{(2\pi)^3 2 E_k}
\int \frac{d^3 k'}{(2\pi)^3 2 E_{k'}}
\nonumber\\
&\times& (2\pi)^4 \delta^4(p + p' - k - k') \nonumber\\
&\times& \Bigl(
-|{\cal M}_{\chi {\cal B} \to \chi {\cal B}}|^2 
f_\chi(\vec{p}) f^{\rm eq}_{\cal B}(\vec{k}) (1 \pm f_\chi(\vec{p'}))\nonumber\\
&\times&(1 \pm f^{\rm eq}_{\cal B}(\vec{k'})) 
+|{\cal M}_{\chi {\cal B} \to \chi {\cal B}}|^2 
f_\chi(\vec{p'}) f^{\rm eq}_{\cal B}(\vec{k'}) \nonumber\\
&\times&(1 \pm f_\chi(\vec{p})) (1 \pm f^{\rm eq}_{\cal B}(\vec{k}))
\Bigr),
\end{eqnarray}
where ${\cal B}$ and ${\cal B}'$ stand for particles in the thermal bath such as ALPs and photons, 
$g_\chi$ is the number of internal degrees of freedom of DM,
and $f_{\cal B}^{\rm eq}$ is given by the Fermi-Dirac or Bose-Einstein distribution depending on the spin of ${\cal B}$.
The summation should be taken for all the internal degrees of freedom
for all the particles.
For the nonrelativistic DM, $C_{\rm el.}$ can be simplified as the Fokker-Planck operator~\cite{Binder:2016pnr,Bertschinger:2006nq, Bringmann:2006mu, Bringmann:2009vf}: 
\begin{align}
\label{Eq:FP}
C_\mathrm{el} 
&\simeq
\frac{E}{2} \gamma(T)
{\Bigg [}
T E\partial_p^2 \!+ \left(  \frac{2 T E}{p} \!+\! p \!+\! \frac{T p}{E}\right) \partial_p + 3
{\Bigg ]}f_{\chi}.
\end{align}
where $T$ is the temperature of the thermal bath and $\partial_p\equiv\partial/\partial p$. 
In the above equation, the momentum transfer rate $\gamma(T)$ is given by (see also 
Ref.~\cite{Gondolo:2012vh}) 
\begin{eqnarray}
\label{eq:gamma}
\gamma&=& \frac{1}{3 g_{\chi} m_{\chi} T} \! \int \! \frac{\text{d}^3 k}{(2\pi)^3} f_{\cal B}^{\pm}(E_k)\left[1\!\mp\! f_{\cal B}^{\pm}(E_k)\right] \nonumber\\
&\times&\int\limits^0_{-4 k_\mathrm{cm}^2} \! \! \!  \text{d}t (-t) \frac{\text{d}\sigma}{\text{d}t} v\,,
\end{eqnarray}
%
where $f_{\cal B}^{\pm}(E_k)$ is Fermi-Dirac or Bose-Einstein distribution of bath particle, $v$ is the relative velocity and $t$ is the Mandelstam variable. 
The differential cross section can be expressed as $({\text{d}\sigma}/{\text{d}t}) v$ $ \equiv$ $|\mathcal{M}|^2_{\chi {\cal B}\leftrightarrow\chi {\cal B}}$ /$(64 \pi {k} E_k m_\chi^2)$, and 
$k_{\rm cm}^2$ is given by 
\begin{eqnarray}
 k_{\rm cm}^2 =\frac{m_\chi^2 ( E_k^2 - m_{\cal B}^2)}{m_\chi^2 + m_{\cal B}^2 + 2 m_\chi E_k}.
\end{eqnarray}
Here $E_k$ is the energy of  heat bath particle ${\cal B}$.
Note that $k_{\rm cm}^2 \neq  E_k^2 - m_{\cal B}^2 = |\vec{k}|^2$. The integrals of the amplitudes of the elastic scattering processes are presented in Appendix~\ref{appendix:cross-sections}.


During the chemical decoupling, the scattering processes may not be frequent enough to maintain the kinetic equilibrium, which means that DM particles own a different temperature $T_\chi$ from the thermal plasma in their following evolution.  
A common definition of the DM temperature is 
\begin{eqnarray}
 T_\chi (T)
&=&
\frac{g_\chi}{3 n_\chi}\int \frac{d^3 p}{(2\pi)^3} \frac{\vec{p}^2}{E} f_\chi (\vec{p})\equiv
\frac{s^{2/3}}{m_\chi} y,
\end{eqnarray}
which is also a function of the thermal bath temperature $T$. In this definition $n_\chi$ is the number density of the DM,
and $s$ is the entropy density.  Here $y$ is a dimensionless version in analogy to the DM yield $Y (=n_\chi/s)$.

To reach a suitable description of the DM temperature evolution and then explore the early kinetic decoupling (EKD) effect on the chemical decoupling process, we should 
consider the second moment of $f_\chi$ as a dynamical degree of freedom. 
By integrating Eq.~(\ref{Eq:boltzmann}) with $g_\chi \int \frac{d^3 p}{(2\pi)^3} \frac{1}{E}$
and
$g_\chi \int \frac{d^3 p}{(2\pi)^3} \frac{1}{E} \frac{\vec{p}^2}{E^2}$, one obtains the zeroth and second moments of the Boltzmann equation, respectively. 
This leads to a relatively simple coupled system of Boltzmann differential equations, 
\begin{eqnarray}
\frac{Y'}{Y} &=& \frac{s Y}{x \tilde H}\left[
\frac{Y_{\rm eq}^2}{Y^2} \left\langle \sigma v\right\rangle_T- \left\langle \sigma v\right\rangle_{T_\chi}
\right]\,, \label{Yfinalfinal}\\
\frac{y'}{y} &=&   \frac{1}{x\tilde H}\langle C_{\text{el}} \rangle_2
+\frac{sY}{x\tilde H}\left[
\left\langle \sigma v\right\rangle_{T_\chi}-\left\langle \sigma v\right\rangle_{2,T_\chi}
\right]
+\frac{sY}{x\tilde H} \nonumber\\ 
&\times&\frac{Y_{\rm eq}^2}{Y^2}\left[
\frac{y_{{\rm eq}}}{y}\left\langle \sigma v\right\rangle_{2,T}\!-\!\left\langle \sigma v\right\rangle_T
\right]
+2(1-w)\frac{H}{x\tilde H}\,,\label{eq:cbe}
\end{eqnarray}
where $x$ is defined as usual $x = m_\chi/T$, 
$Y_{\rm eq}(x)\equiv n_{\rm eq}(T)/s$, and 
\begin{eqnarray}
\tilde H\equiv H/\left[1+ (1/3)d(\log g^s_{\rm eff})/d(\log T)\right], 
\end{eqnarray}
with $g^s_{\rm eff}$ being the entropy degrees of freedom of the 
background plasma. 
$w(T_\chi)\equiv 1-{\langle p^4/E^3 \rangle_{T_\chi}}/({6T_\chi})$, with $\langle p^4/E^3\rangle
=\frac{g_\chi}{n_\chi^{\rm eq}(T_\chi)}\int \frac{d^3p}{(2\pi)^3} {\left( \vec{p} \cdot \vec{p} \right)^2}/{E^3} e^{- \frac{E}{T_\chi}}$. 
Note that the elastic scattering term given in Eq.~(\ref{Eq:FP}) does not contribute to 
the zeroth moment term. This is a natural consequence because the elastic scattering processes do not change the number density of DM.

The above compact form of the differential equations contains the following thermally averaged cross sections, 
\begin{align}
\langle C_{\text{el}} \rangle_2 \equiv \frac{g_\chi}{3 n T_{\chi}} \int \frac{\text{d}^3 p}{(2\pi)^3} \frac{p^2}{E^2} C_{\text{el}}\;,
\end{align}
\begin{eqnarray}
\VEV{\sigma v}_{T_\chi}
&\equiv&
\frac{g_\chi^2}{(n_\chi^{\rm eq})^2}
\int \frac{d^3 p}{(2\pi)^3} 
\int \frac{d^3 q}{(2\pi)^3} 
\left( \sigma v \right)_{\chi \chi \to {\cal B} {\cal B}'}\nonumber\\
&\times&
f_\chi^{\rm eq}(\vec{p}, T_\chi)
f_\chi^{\rm eq}(\vec{q}, T_\chi). 
\end{eqnarray}
The thermal average $\langle \sigma v\rangle_{2,T}$ is a variant of 
the commonly used thermal average $\langle \sigma v\rangle_T$, and is explicitly stated in 
Ref.~\cite{Binder:2017rgn} and introduced as 
\begin{eqnarray}
\VEV{\sigma v}_{2, T_\chi}
&=&
\frac{g_\chi^2}{(n_\chi^{\rm eq})^2 T_\chi}
\int \frac{d^3 p}{(2\pi)^3} 
\int \frac{d^3 q}{(2\pi)^3} 
\frac{\vec{p} \cdot \vec{p} }{3E} 
\left( \sigma v \right)_{\chi \chi \to {\cal B} {\cal B}'}\nonumber\\
&\times&
f_\chi^{\rm eq}(\vec{p}, T_\chi)
f_\chi^{\rm eq}(\vec{q}, T_\chi). 
\end{eqnarray}
For 
$\VEV{\sigma v}_{T} $
and 
$\VEV{\sigma v}_{2, T} $,
we replace $T_\chi$ by $T$ in 
$\VEV{\sigma v}_{T_\chi} $
and 
$\VEV{\sigma v}_{2, T_\chi} $, respectively.
Moreover, $n_\chi^{eq}(T_\chi)$ is given by
\begin{eqnarray}
 n_\chi^{\rm eq}(T_\chi) = g_\chi \int \frac{d^3 p}{(2\pi)^3} f_\chi^{\rm eq}(\vec{p}, T_\chi)
= g_\chi \int \frac{d^3 p}{(2\pi)^3} e^{- \frac{E_p}{T_\chi}}.
\end{eqnarray}

\subsection{EKD effect in the resonance mechanism}

\begin{figure}[h]
 \centering
  \includegraphics[width=0.475\textwidth]{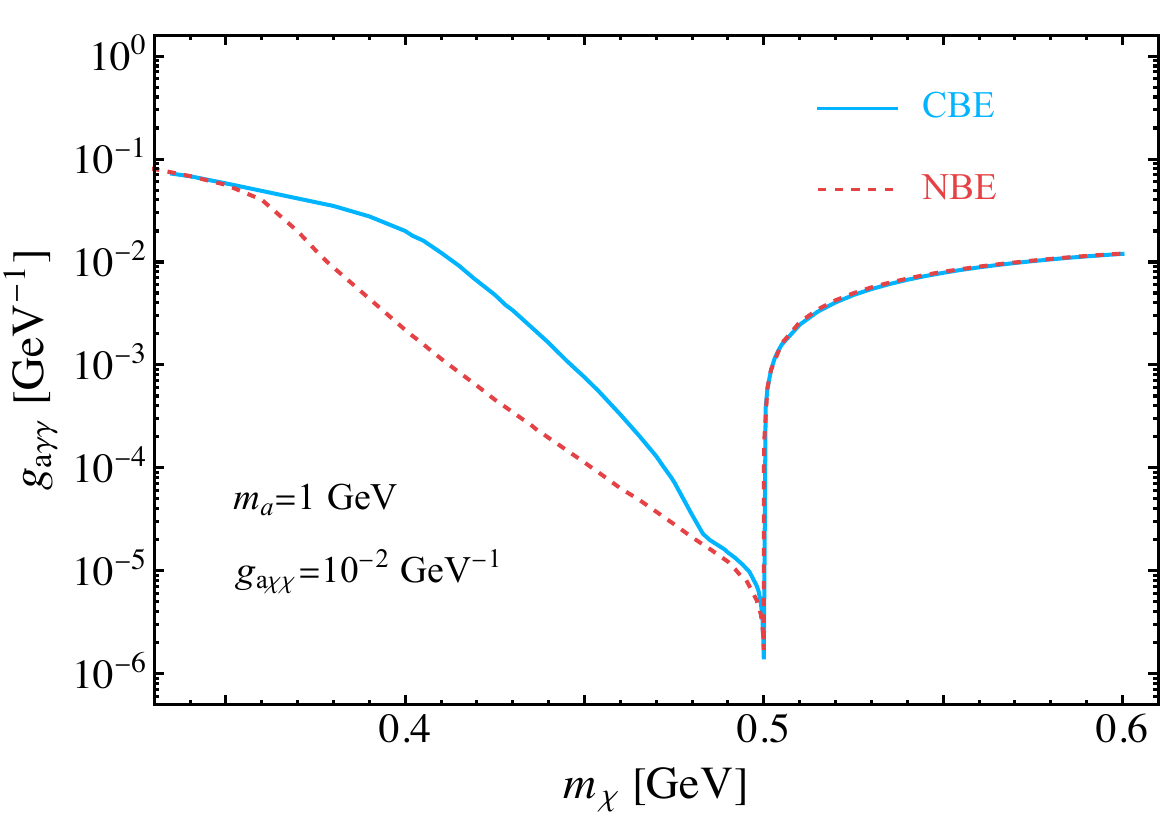}  
  \caption{The required value of the coupling $g_{a\gamma\gamma}$ to achieve the correct relic density is shown in the figure. The red dashed line represents the results using the NBE method, which relies on the assumption of kinetic equilibrium throughout the entire freeze-out epoch. The blue line corresponds to the results derived from the CBE method. For illustration, we adopt $m_a=1~\rm GeV$ and $g_{a\chi\chi}=10^{-2}~\rm GeV^{-1}$.}
  \label{fig:mchigagg}
\end{figure}

Initially, we compute the relic density using both the standard treatment and the CBE treatment. 
We numerically solve Eqs.~(\ref{Yfinalfinal}) and~(\ref{eq:cbe}) for a given set of parameters ($m_\chi, g_{a\gamma\gamma}$) and determine the resulting asymptotic value of $Y_0$. For illustration, we fix $m_a=1~\rm GeV$ and $g_{a\chi\chi}=10^{-2}~\rm GeV^{-1}$. 
The viable parameter space is determined by the matching the observed relic density reported by the Planck Collaboration, which is $\Omega h^2 = 0.120\pm 0.001$~\cite{Planck:2018vyg}. The red dashed line in Fig.~\ref{fig:mchigagg}, corresponding to the NBE method, 
shows the contour in this plane that yields the correct relic density. 
In the same figure, we also display the required value of $g_{a\gamma\gamma}$ obtained by solving the coupled system of Boltzmann equations, represented by the blue solid curve. 
Outside the resonance region, the coupled Boltzmann equations produce results identical to those of the standard approach, indicating that kinetic decoupling occurs much later than chemical decoupling and that the assumption of local thermal equilibrium during chemical freeze-out is satisfied. 
However, for DM masses within the resonance region, the CBE method yields significantly different results, suggesting that this assumption is violated. 
In the resonance region,  the momentum transfer rate $\gamma(T)$ is strongly suppressed compared to the annihilation rate. This suppression arises because the small coupling $g_{a\gamma\gamma}$ required to satisfy the relic density condition does not benefit from a corresponding resonant enhancement of $\gamma(T)$, leading to the early decoupling of kinetic equilibrium.

\begin{figure}[h]
  \centering
  \includegraphics[width=0.475\textwidth]{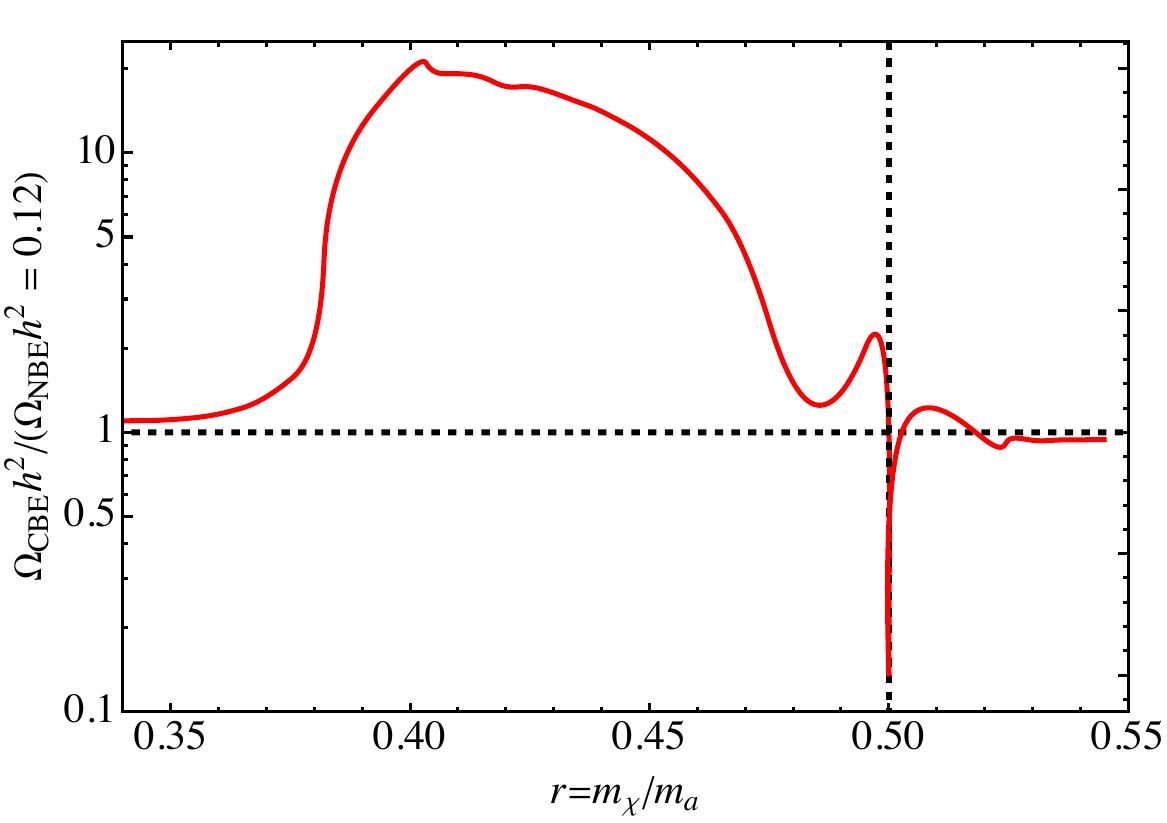} 
  \caption{The relic density ratio between CBE and NBE methods, where we require the parameters to match the observed abundance in the NBE approach, i.e. $\Omega_{\rm NBE} h^2=0.12$. Here we set $m_a=1$ GeV. 
  }
  \label{fig:c/nBE}
\end{figure}

\begin{figure}[!htbp]
  \centering
\includegraphics[width=0.47\textwidth]{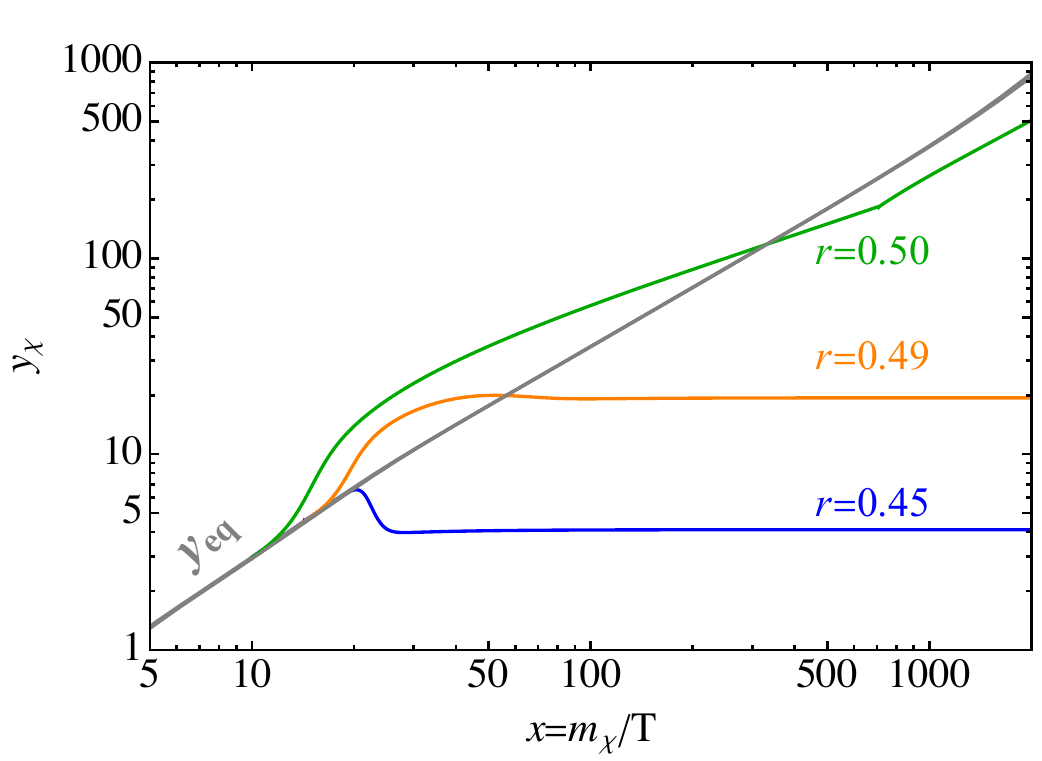}  
  \caption{The evolution of the DM `temperature' $y_{\chi}$ with respective to $x=m_{\chi}/T$. The gray line represents the temperature of the thermal bath. The blue, orange, and green curves correspond to $r\equiv m_{\chi}/m_a=0.45,~0.49$, and $0.5$, respectively. Here we take $m_a=1 \rm GeV$ for illustration. 
  }
  \label{fig:temperature}
\end{figure}

\begin{figure}[!htbp]
  \centering
  \includegraphics[width=0.235\textwidth]{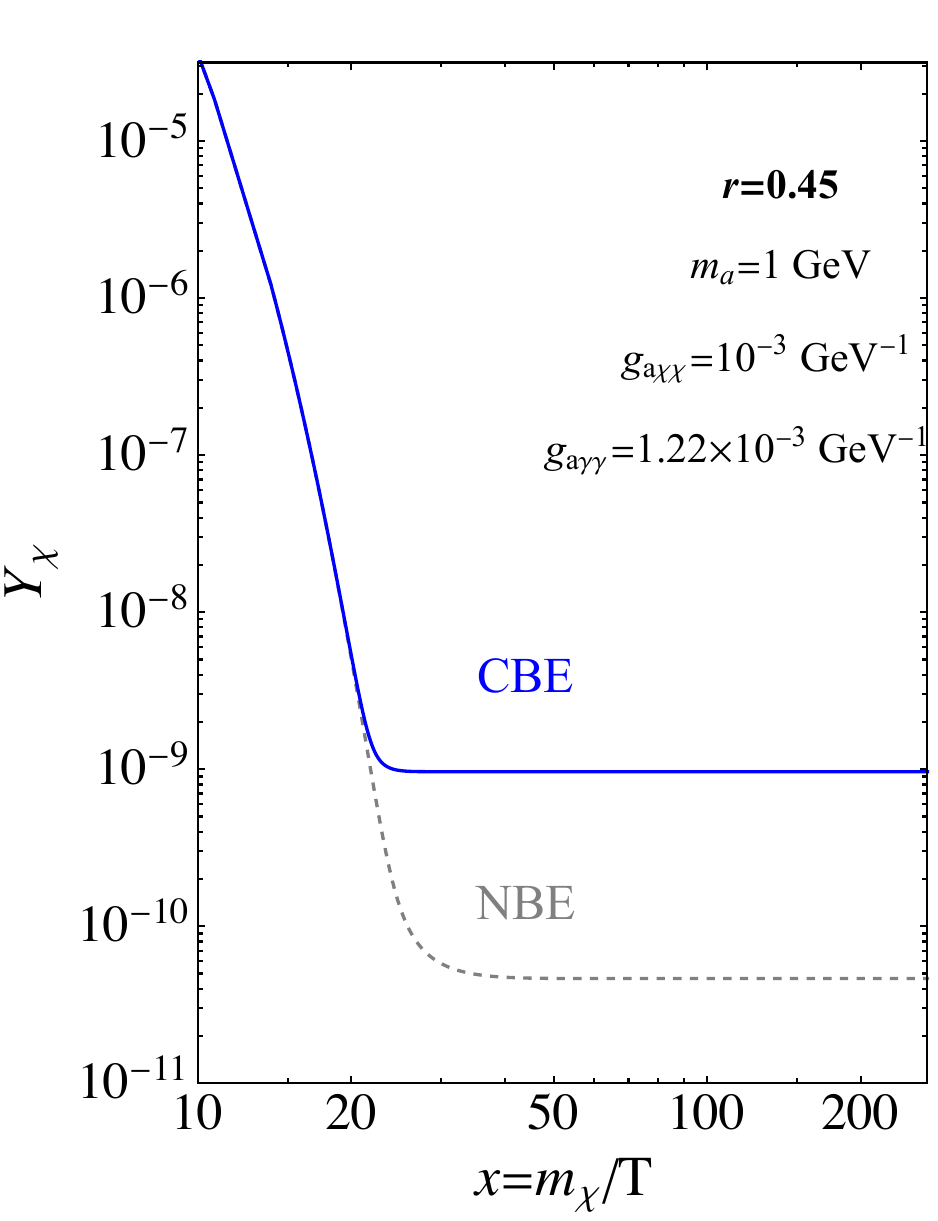}  
\includegraphics[width=0.235\textwidth]{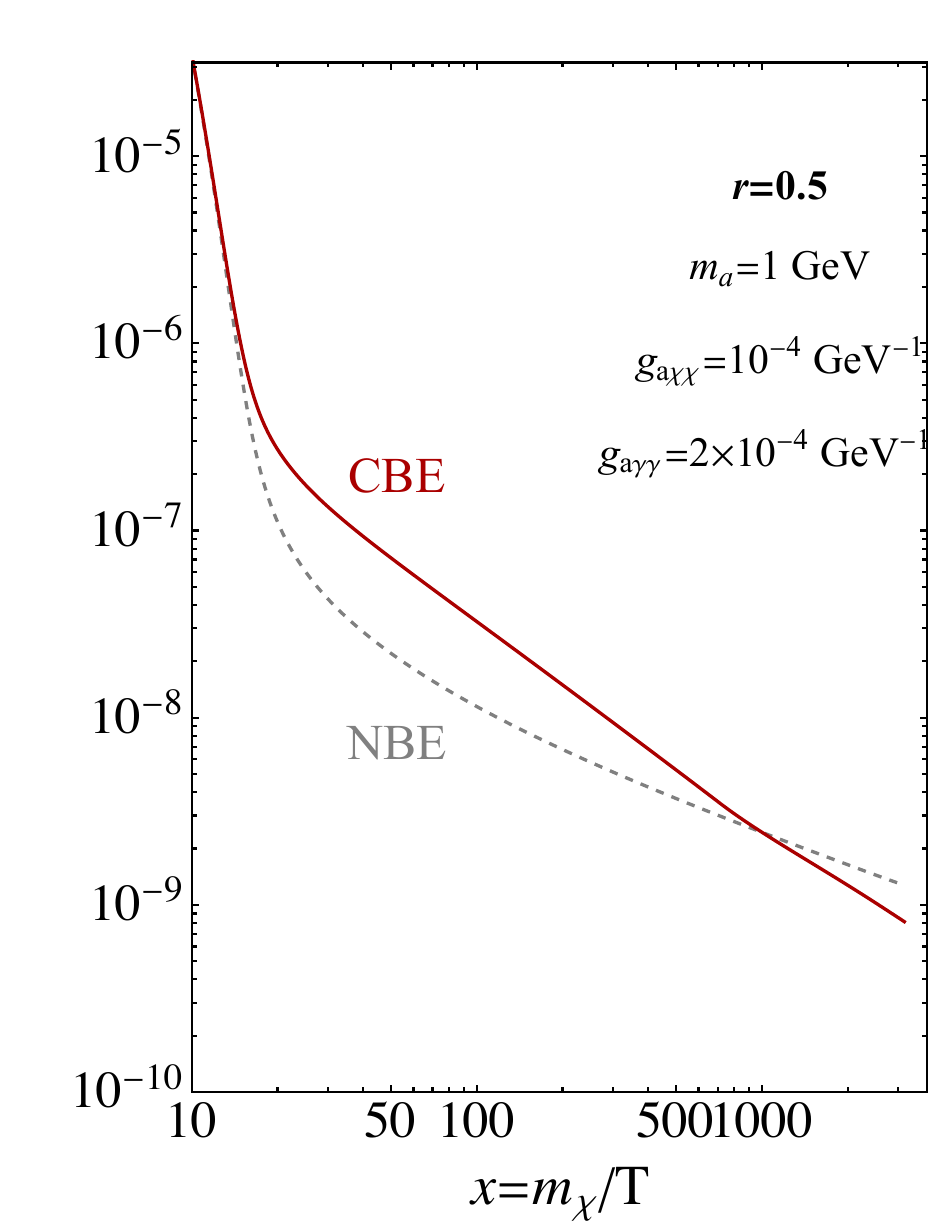}  
\caption{
  The evolution of DM abundance $Y_\chi$ with respect to $x=m_{\chi}/T$. Left panel: The blue line depicts the DM yields in the CBE approach, with benchmark parameters labeled in the plot. The gray dashed line represents the NBE approach. Right panel: The red line shows the DM yields in the CBE approach for a different set of benchmark parameters. The gray dashed line corresponds to the yield in the NBE approach. We observe significantly different evolutionary trajectories for varying mass ratios $r\equiv m_{\chi}/m_a$.}
  \label{fig:Yield}
\end{figure}

To highlight the significance of the improved treatment of decoupling dynamics near the resonance region, we plot in Fig.~\ref{fig:c/nBE} the ratio of the relic density obtained from the CBE method to that from the standard NBE approach. The parameters used here satisfy the requirement $\Omega_{\rm NBE}h^2 = 0.12$. 
As shown in the figure, the CBE results differ significantly from the NBE results for the same parameters. First, as demonstrated by the benchmark points with $r<0.37$ and $r>0.55$, for DM masses sufficiently far from the resonance, the phase-space distribution remains nearly Maxwellian in shape. For these cases, we observe, as expected, excellent agreement in the evolution of $Y(x)$ and $y(x)$ between the CBE and the NBE approaches. However, in the resonance region, the ratio $\Omega_{\rm CBE}/\Omega_{\rm NBE}$ exhibits a complex structure. For instance, $\Omega_{\rm CBE}/\Omega_{\rm NBE} \simeq 20$ when $r\simeq 0.41$. This behavior can be understood by dividing the region into three distinct parts: 
\begin{itemize}

\item Sub-resonance region ($0.37\lesssim r\lesssim 0.48$)

 EKD happens in this region, the DM phase-space density $f_\chi$ begins to deviate from the equilibrium value of the thermal bath, requiring higher momenta to reach the resonance.

Consider an illustrative case where $r=0.45$. In this scenario, resonant annihilation depletes $f_\chi$ for momenta above the peak of the distribution~\cite{Binder:2017rgn}. This leads to a relative reduction in $f_\chi$ compared to a thermal distribution at these momenta, thereby decreasing the DM velocity dispersion, often referred to as the ``temperature''. This effect is clearly visible in Fig.~\ref{fig:temperature}, where the evolution of $y_\chi$ shows a decline, as indicated by the blue curve. 

Moreover, this behavior can be directly understood from Eq.~(\ref{eq:cbe}). Specifically, in the sub-resonant regime, we find that $\langle \sigma v\rangle_{T_\chi}<\langle \sigma v\rangle_{2,T_\chi}$. This drives $y_\chi$ to smaller values after decoupling. Additionally, the depletion in $f_\chi$ reduces $\langle \sigma v\rangle_{T_\chi}$, leading to an earlier chemical decoupling. As a result, the relic density increases, as demonstrated in the left panel of Fig.~\ref{fig:Yield}.

\item  Resonance region ($0.48\lesssim r\lesssim 0.5$)

When entering the resonant region, DM annihilation begins to deplete $f_\chi$ below the peak of the would-be Maxwellian distribution. As a result, once equilibrium is no longer maintained (i.e. EKD), the depletion causes an increase in the velocity dispersion, in contrast to the decrease observed in the sub-resonant regime. This phenomenon is clearly illustrated by the orange and green curves in Fig.~\ref{fig:temperature}.

At the exact resonance point, the residual DM particles with higher velocities contribute to an enhanced cross section $\langle \sigma v\rangle$. This increase in $\langle \sigma v\rangle$ reduces the DM abundance, as shown in the right panel of Fig.~\ref{fig:Yield}. This mechanism explains the formation of the narrow dip observed near the resonance position, as depicted in Fig.~\ref{fig:c/nBE}. 

\item Super-resonance region ($0.5\lesssim r\lesssim 0.55$)

In the super-resonant regime, characterized by $r\gtrsim 0.5$, the condition $\sqrt{s}\gtrsim m_a$ is inherently satisfied. Consequently, a resonantly enhanced annihilation rate can only occur within a highly confined region of phase space, where the relative DM momenta are nearly zero. This restriction implies that the influence of the resonance diminishes rapidly, rendering its effect practically negligible in this regime.
\end{itemize}

\section{Combined Results}
\label{sec:results}

The determination of the DM relic density utilizing the CBE method enables a precise delineation of the parameter space, as discussed in the previous section. This parameter space represents a promising regime for further exploration, particularly in light of the observed excess in the $B^\pm \to K^\pm \nu\bar{\nu}$ decay mode. In this section, we will present the combined results that simultaneously account for the observed DM relic abundance and the $B^\pm \to K^\pm \nu\bar{\nu}$ excess.

\subsection{Displaced diphoton scenario}

As discussed in Sec.~\ref{sec:ALPs},  ALPs predominantly undergo decay into diphotons in various scenarios, including DM $p$-wave annihilation, forbidden annihilation, and $s$-wave annihilation with certain coupling configurations (refer to Fig.~\ref{fig:BR}). In these scenarios, ALPs are necessitated to possess low masses, and the coupling $g_{a\gamma\gamma}$ must be sufficiently small to guarantee an ALP lifetime that permits them to traverse beyond detectors at the Belle II. This results in a signature characterized by missing energy.

To explain the observed $B^\pm \rightarrow K^\pm \nu\bar{\nu}$ excess at the Belle II, we should determine the number of signal events that can generate invisible ALP signature at the detector for this scenario. This can be expressed as follows:
\begin{eqnarray}
    N(B^\pm\to K^\pm \slashed{E} )&=&N_B\times Br(B^\pm \rightarrow K^\pm a)\nonumber\\
    &\times& Br(a\to\gamma\gamma)\times P_a(R).
    \label{eq:NB}
\end{eqnarray}
Here, $N_B$ denotes the number of B mesons produced at the Belle II. The probability that an ALP is detected at a distance $R$ from the interaction point is given by
\begin{eqnarray}
P_a(R)=\exp(-\frac{R}{L_a}), \quad~L_a=\gamma\beta c \tau_a,
\label{eq:Pa}
\end{eqnarray}
where $\gamma$ represents the Lorentz factor, $\beta$ denotes the velocity of the ALP, and $\tau_a\simeq1/\Gamma_{a\to\gamma\gamma}$  denotes the proper decay time of the ALP. In this scenario, $m_a\ll m_{B^\pm}$, and therefore the boost factor can be estimated as $\gamma\beta\simeq m_{B^\pm}/(2m_a)$.

The structure and geometry sizes of detectors are crucial in determining whether the signature of an ALP is a displaced diphoton or missing energy in Eq.~(\ref{eq:NB}). At the Belle II detector, for 
$R > 150$ cm in radius from the interaction point, only the $K_L$-Muon (KLM) detector is covered \cite{Duerr:2020muu}. The KLM detector is designed to detect $K_L^0$ mesons, neutrons, and muons, and has negligible efficiency for detecting a photon pair \cite{Belle-II:2018jsg}. 
Therefore, we use $R > 150$ cm in Eq.~(\ref{eq:Pa}) as a simple criterion for the long-lived ALP to be invisible in the Belle II  experiment.

\begin{figure}[!htbp]
  \centering
  \includegraphics[width=0.48\textwidth]{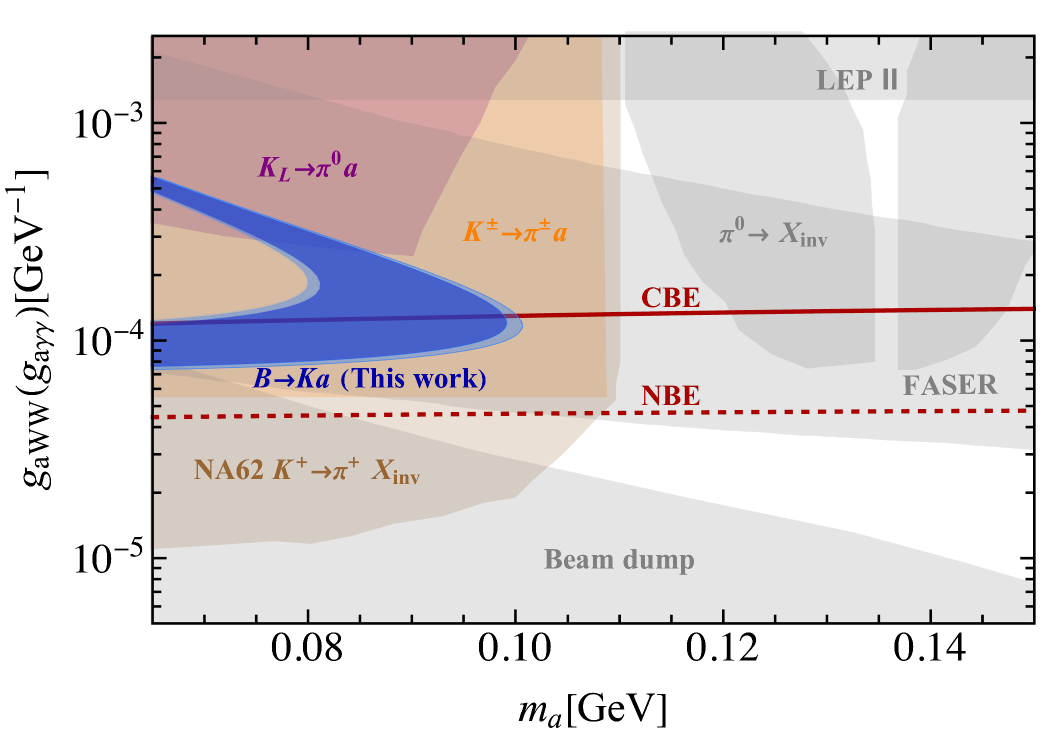}  
  \caption{
Signal region and experimental constraints in the ALP parameter space, assuming ALP decays into displaced diphotons. We apply the Belle II search for \( B \to K\nu \bar{\nu} \)~\cite{Belle-II:2023esi} to constrain the decay \( B \to Ka \) (blue region). 
The parameter regions excluded by existing bounds: 
$K^\pm \rightarrow\pi ^\pm X_{inv}$ from NA62 experiment~\cite{NA62:2021zjw} (brown region), $\pi ^0\rightarrow X_{inv}$ from NA62 experiment~\cite{NA62:2020pwi}, the LEP experiment~\cite{Jaeckel:2015jla}, 
and the searches of the long-lived particle decaying into a pair of photons by the FASER experiment \cite{FASER:2024bbl} (gray regions).  
Also the constraints from $K_L \rightarrow\pi ^0a,~ K^\pm\rightarrow \pi^\pm a$, and other Beam Dump experiments \cite{Izaguirre:2016dfi} are included. 
The solid red line represents the required value of the coupling $g_{a\gamma\gamma}$ to achieve the correct relic density in the CBE approach, while the dashed red line represents the corresponding value in the NBE approach. Here, we take $r = 0.47$ and $g_{a\chi\chi} = 10^{-3} \, \text{GeV}^{-1}$ as a benchmark point.}
  \label{fig:ALPgamma}
\end{figure}

To estimate the bounds for the blind zone of $100 ~{\rm MeV} < m_a < 175~\rm MeV$ and for even lighter ALPs, we employ Eq.~(\ref{eq:NB}) in conjunction with the Belle II (2023) measurement, ${\rm Br}(B^{\pm} \rightarrow K^{\pm} \nu \bar{\nu})=(2.3 \pm 0.7) \times 10^{-5}$~\cite{Belle-II:2023esi}. The resulting bounds are illustrated as the blue bulk in Fig.~\ref{fig:ALPgamma}. The shape of this new bound is influenced by the ALP production rate and its decay length in the lab frame. As the parameters $g_{aWW}$ and $m_a$ decrease, the decay length of the ALP in the lab frame increases significantly, but the production rate from $B^\pm \rightarrow K^\pm a$ decreases considerably.  
The lighter blue region arises from uncertainties in the form factor.

For further investigation, we incorporate existing constraints such as those derived from the  $K^\pm \to \pi^\pm + \rm X_{inv.}$ search conducted by the NA62 experiment~\cite{NA62:2021zjw} (represented by the brown area) and the $\pi ^0\rightarrow X_{inv}$ search by the same experiment~\cite{NA62:2020pwi} (depicted as the light gray region). Additionally, we present constraints stemming from searches of $K_L \rightarrow\pi ^0a$ and $K^\pm\rightarrow \pi^\pm a$ processes, the LEP experiment~\cite{Jaeckel:2015jla}, and various Beam Dump experiments~\cite{Izaguirre:2016dfi}. Furthermore, we include the constraint imposed by the FASER experiment's search for long-lived particles decaying into a pair of photons~\cite{FASER:2024bbl} (indicated by the gray area). Notably, our newly delineated parameter space in this specific scenario completely overlaps with the exclusion limits, suggesting that the scenario involving displaced diphoton final states of ALPs is unable to account for the observed excess.

Since we have set \( g_{a\gamma\gamma} = g_{aWW} \), we are able to present the parameter space corresponding to the DM relic density within the same plot. We depict this parameter space using a solid (dashed) red curve to signify the results derived from the CBE (NBE) approach. For the purpose of illustration, we adopt \( r = 0.47 \) and \( g_{a\chi\chi} = 10^{-3}\,\text{GeV}^{-1} \) as our benchmark values.  It is evident that the CBE results significantly diverge from those of the NBE, highlighting the need for a more precise treatment in our analysis.

\subsection{Missing energy scenario}

\begin{figure}[!htbp]
\centering  
\includegraphics[width=0.48\textwidth]{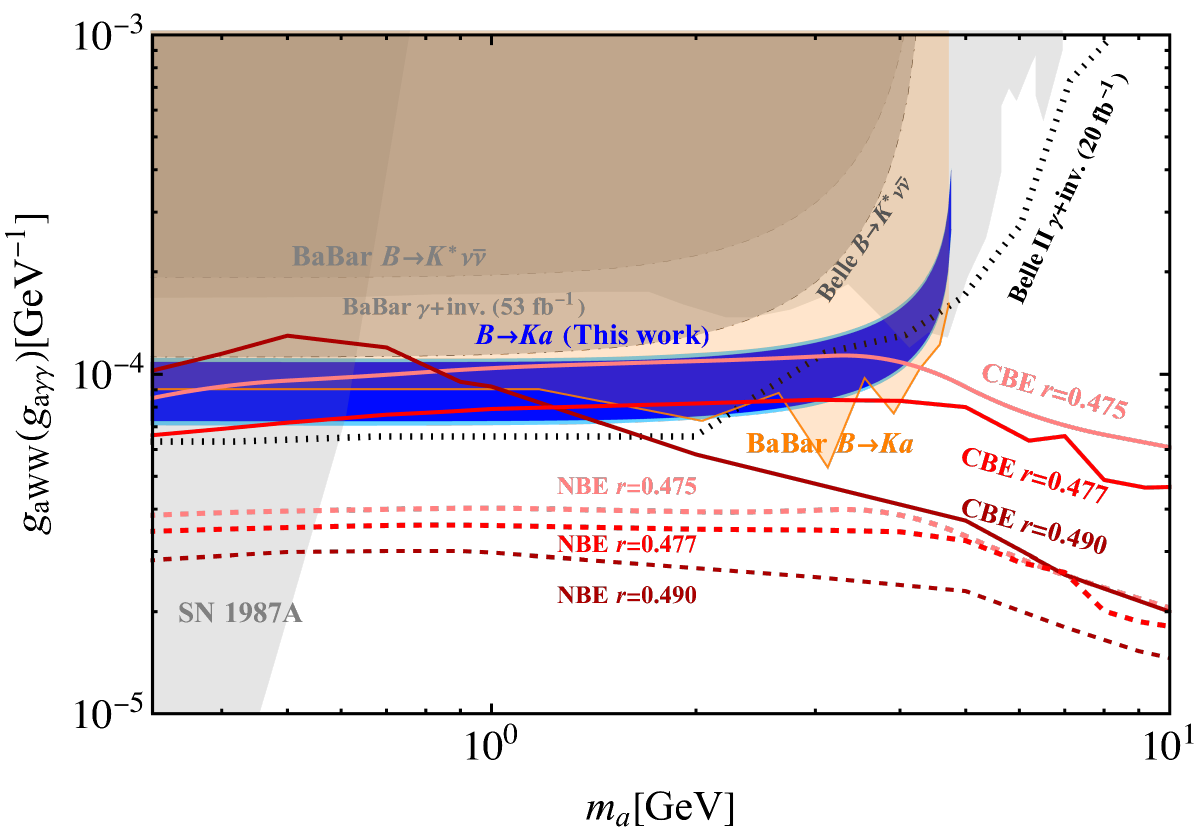} 
\caption{
Signal region and experimental constraints in the ALP parameter space assuming ALP decays to a pair of DM particles. The blue bulk represents the parameter space for $B \to Ka$ determined by the $B^\pm \rightarrow K^\pm \nu\bar{\nu}$ measurement from Belle II~\cite{Belle-II:2023esi}. 
We show existing constraints from SN 1987A \cite{Dolan:2017osp}, 
BaBar $\gamma+inv.~(53~\rm fb^{-1})$  \cite{BaBar:2017tiz}, BaBar $B\to K^* \nu\bar{\nu}$\cite{BaBar:2013npw}  and Belle  $B\to K^* \nu\bar{\nu}$ \cite{Belle:2017oht} searches  (gray regions).  
The constraints are also derived from limits on ALP production in the decay of $B\rightarrow Ka$ by BaBar data \cite{Izaguirre:2016dfi} (orange).  
Additionally, the expected sensitivity from Belle II $\gamma +inv~(20 \rm ~fb^{-1})$ data \cite{Dolan:2017osp} is shown by black dotted curve. 
The solid and dashed  red (and light-red) curves have the same meaning as in Fig.~\ref{fig:ALPgamma}.   
}
\label{fig:ALPDM}
\end{figure}

When the invisible decay channel of ALPs dominates in the DM $s$-wave annihilation scenario, the lifetime of ALPs becomes less of a concern, which directly accounts for this excess. The number of signal events can be expressed as follows:
\begin{eqnarray}
    N(B^\pm\to K^\pm \slashed{E} )&=&N_B\times Br(B^\pm \to K^\pm a) \nonumber\\
    &\times&Br(a\to {\rm inv.}).
    \label{eq:NBDM}
\end{eqnarray}
Using the $B^\pm \rightarrow K^\pm \nu\bar{\nu}$ measurement from Belle II~\cite{Belle-II:2023esi}, we are able to ascertain the feasible parameter region for the decay of an ALP into a pair of DM particles, within the mass range where $m_a$ is less than $m_{B^\pm}-m_{K^\pm}$. The outcomes of this analysis are represented in Fig.~\ref{fig:ALPDM}, highlighted in the blue region. Our findings indicate that the $B^\pm \rightarrow K^\pm \nu\bar{\nu}$ excess prefers a coupling strength $g_{aWW}$ within the interval of $(7.3\times 10^{-5} \sim 1.1 \times 10^{-4})~ \rm GeV^{-1}$ across the majority of the $m_a$ spectrum.

For comparative analysis, the plot also incorporates existing constraints, namely those derived from SN 1987A \cite{Dolan:2017osp}, 
BaBar $\gamma+inv.$ (with $53~\rm fb^{-1})$ \cite{BaBar:2017tiz},  BaBar $B\to K^* \nu\bar{\nu}$ \cite{BaBar:2013npw} and Belle  $B\to K^* \nu\bar{\nu}$ \cite{Belle:2017oht} searches. 
Notably, the constraint imposed by SN 1987A excludes the parameter space where $m_a\lesssim0.6$ GeV. BaBar searches of $B\to K\nu\bar{\nu}$ \cite{BaBar:2013npw} put strict constraints on the parameter space of ALP production by process $B\to K a$ as shown in the orange region \cite{Izaguirre:2016dfi}. 
Consequently, there remains limited room to accommodate the excess events. 
Of particular interest, we also include the expected sensitivity from Belle II searches of final state with $\gamma +inv~(20 \rm ~fb^{-1})$  \cite{Dolan:2017osp}, showing that the most parameter space will be tested.

Regarding the DM relic density in this scenario, the parameter space is already subject to stringent constraints from CMB observations and indirect detection experiments. As a result, the DM production mechanism is strongly influenced by the resonance effects. 
Fig.~\ref{fig:ALPDM} displays six curves (shown in (light-) red): the solid curves derived using the CBE method and the dashed ones using the NBE method, both of which satisfy the DM relic density criterion. For clarity, specific parameter combinations are selected: $(r,~g_{a\chi\chi})=(0.475,~10^{-3}~ \rm GeV^{-1})$, $(0.477,~10^{-3}~ \rm GeV^{-1})$, and $(0.490,~1.04\times10^{-5} ~\rm GeV^{-1}$). 
Notably, the CBE curves align closely with the blue region over a substantial range. In contrast, the NBE results deviate significantly from the CBE curves and fail to intersect with the viable region delineated by the Belle II measurement within this framework. Consequently, we conclude that the CBE method yields distinct and more accurate results compared to the traditional approach.

\begin{table}[h]

\centering
\begin{tabular}{c c c c c}
\hline
BP & $m_a$ (GeV) &  $g_{a\gamma\gamma}$ (GeV$^{-1}$) & $g_{a\chi\chi}$ (GeV$^{-1}$) \\
\hline
\multirow{2}{*}{1} & \multirow{2}{*}{$0.75$} & Lower Bound:  $7.13 \times 10^{-5}$ & $7.73 \times 10^{-3}$  \\
                           &                           & Upper Bound: $8.95 \times 10^{-5}$ & $7.12\times 10^{-5}$  \\
\hline
\multirow{2}{*}{2} & \multirow{2}{*}{$2.65$} & Lower Bound: $7.81 \times 10^{-5}$ & $2.18 \times 10^{-3}$  \\
                           &                           & Upper Bound: $8.73 \times 10^{-5}$ & $1.40 \times 10^{-4}$  \\
\hline
\multirow{2}{*}{3} & \multirow{2}{*}{$3.55$} & Lower Bound: $9.0\times 10^{-5}$ & $1.25 \times 10^{-4}$  \\
                           &                           & Upper Bound: $9.60 \times 10^{-5}$ & $8.86 \times 10^{-5}$  \\
\hline

\end{tabular}
\caption{Viable range of coupling $g_{a\chi\chi}$ for the selected benchmark points (BPs). The table presents the ALP mass points with the allowed range of $g_{a\gamma\gamma}$ selected as benchmarks from the remaining mass intervals after the signal region is excluded by experimental limits. The lower and upper bounds of $g_{a\gamma\gamma}$ are derived from the signal and experimental exclusion regions in Fig. \ref{fig:ALPDM}. The corresponding values of $g_{a\chi\chi}$ are computed where the relic density of DM is equal to 0.12 using the CBE method. Here $r=0.477$ is chosen.}
\label{tab:couplings}
\end{table}

To better interpret the influence of the coupling $g_{a\chi\chi}$, we select ALP mass points from the remaining mass intervals that have not been ruled out by experimental limits within the signal region as benchmarks, as illustrated in Table~\ref{tab:couplings}. For each of these regions, the combination of signal region and experimental exclusions yields both a lower and an upper bound for $g_{a\gamma\gamma}$. We then use the CBE method to derive the corresponding values of $g_{a\chi\chi}$ that ensure the DM has the observed relic density.

\section{Conclusion}
\label{sec:conclusion}

The \(B^\pm \to K^\pm \nu \bar{\nu}\) anomaly observed at Belle II provides a compelling opportunity to explore ALP-mediated dark matter scenarios. Within the resonant annihilation framework (\(m_a \sim 2m_\chi\)), we demonstrate that ALP couplings to both the Standard Model ($g_{aWW} (g_{a\gamma\gamma})$) and dark matter $(g_{a\chi\chi})$ can simultaneously explain the \(B^\pm \to K^\pm \nu \bar{\nu}\) excess and satisfy DM relic density constraints. Key advancements in this work include:
\begin{enumerate}
  
  \item  Distinct ALP decay signatures: Invisible decays (\(a \to \chi\bar{\chi}\)) dominate for \(g_{a\chi\chi} \geq 10^{-4}\, \text{GeV}^{-1}\), producing missing energy signals consistent with the $B^\pm \to K^\pm \nu \bar{\nu}$ excess at the Belle II, while evading SN 1987A and BaBar $B\to K\nu\bar{\nu}$ constraints for \(m_a > 0.6\, \text{GeV}\). Notably, this parameter space can be further explored in the coming years through invisible ALPs searches at the Belle II. In contrast, the displaced diphoton scenario for explaining the $B^\pm \to K^\pm \nu \bar{\nu}$ excess is ruled out by the \(K^\pm \to \pi^\pm a\) searches.  
  
    \item Resonant enhancement and early kinetic decoupling: Utilizing the CBE method, we find the discrepancies between traditional relic density calculations. In resonant regions (\(r = m_\chi/m_a \approx 0.4-0.5\)), EKD effects affect the freeze-out process, amplifying relic densities by up to 20 times compared to standard treatments.

  \item Viable parameter space: The interplay between \(g_{aWW}(g_{a\gamma\gamma}) \in (7.13 \times 10^{-5} - 9.60 \times 10^{-5})\, \text{GeV}^{-1}\) and \(g_{a\chi\chi} \in (7.12\times10^{-5} - 7.73\times 10^{-3})\, \text{GeV}^{-1}\) defines a testable regime for the not excluded ALP mass intervals in the range of \(m_a \in (0.6-4.8)~\text{GeV}\) that can simultaneously explain the $B^\pm \to K^\pm \nu \bar{\nu}$ excess at the Belle II and the observed dark matter relic abundance.
\end{enumerate}

This work establishes the ALP portal as a robust framework for unifying the $B^\pm \to K^\pm \nu \bar{\nu}$ anomaly and thermal dark matter production. Further studies (e.g., B-factories, beam-dump experiments, and astrophysical phenomenology) will provide deeper insights into this scenario.

\section*{Acknowledgments}

We are grateful to  Shu-Yuan Guo for the valuable discussions. 
The work by X.L. is supported by  the Natural Science Foundation of Shandong province under the Grant No. ZR2025MS02 and the Project of Shandong Province Higher Educational Science and Technology Program under Grants No. 2022KJ271. 
Y.L. is supported in part by the National Natural Science
Foundation of China (NNSFC) under the Grants No. 11925506, 12375089, 12435004, and the Natural Science Foundation of Shandong province under the Grant No. ZR2022ZD26. 
C.T.L. is supported by NNSFC under grant No. 12335005 and the Special funds for postdoctoral overseas recruitment, Ministry of
Education of China. 
B.Z. is supported by NNSFC under grant Nos. 12275232 and the Natural Science Foundation of Shandong province under the Grant No. ZR2025QA20.

\appendix
\begin{widetext}
\section{Cross-sections for DM annihilation and scattering processes}
\label{appendix:cross-sections}

The cross-sections for DM annihilation process is
\begin{equation}
    \sigma_{\chi\chi\rightarrow\gamma\gamma}v=\frac{g_{a\chi\chi}^2 g_{a\gamma\gamma}^2 m_{\chi}^2 s^{\frac{5}{2}} \sqrt{s (s-4m_{\chi}^2)}}{2 \pi \sqrt{s-4m_{\chi}^2} (s-2m_{\chi}^2)(m_a^2 \Gamma_a^2 + (m_a^2 -s)^2)},
\end{equation} 
where $s = E^2_{cm}$ is the square of the center-of-mass energy for DM annihilation. 
The decay width of ALP express as 
\begin{equation}
    \Gamma_a ( m_a > 2m_\chi) = \frac{g_{a\chi\chi}^2 m_{\chi}^2 \sqrt{m_a^2-4m_{\chi}^2}}{\pi} + \frac{g_{a\gamma\gamma}^2 m_a^3}{64 \pi}, 
    \label{einstein}
\end{equation}
and 
\begin{equation}
     \Gamma_{a} ( m_a \leq 2m_\chi) = \frac{g_{a\gamma\gamma}^2 m_a^3}{64 \pi},
     \label{eins}
\end{equation}
depending on the mass relations of DM and ALP. 

For relevant scattering processes we have 
\begin{eqnarray}
\int^0_{-4k^2_{\rm cm}} dt (-t) |\mathcal{M}_{\chi\gamma\rightarrow\chi\gamma}|^2&=& 8 g_{a\chi\chi}^2 g_{a\gamma\gamma}^2 m_{\chi}^2 \Bigg(4 m_a^6 \left( \log\left(m_a^2 \right) - \log\left(\frac{4 k^2 m_{\chi}}{m_{\chi} + 2 \omega} + m_a^2 \right) \right) +\frac{4}{3} k^2 m_\chi \notag \\
&\times&  \left( 
\frac{16 k^4 m_{\chi}^2}{(m_\chi + 2\omega)^3} - \frac{12 k^2 m_a^2 m_\chi}{(m_\chi + 2\omega)^2} + \frac{3 m_a^6}{4 k^2 m_\chi + m_a^2 (m_\chi + 2\omega)} + \frac{9 m_a^4}{m_\chi + 2 \omega} 
\right) \Bigg),
\end{eqnarray}
and
\begin{eqnarray}
\int^0_{-4k^2_{\rm cm}} dt (-t) |\mathcal{M}_{\chi a\rightarrow\chi a}|^2&=-\frac{1}{(m_a^2 + 2 m_{\chi} \omega)^2} 256 g_{a \chi}^4 m_{\chi}^2 \Bigg( 
-\frac{1}{3} \left((m_a^2 + 2 m_{\chi} \omega) (m_a^2 + m_{\chi} (m_{\chi} + 2 \omega)) \right. \notag \\
&\quad \left. \times \left(\frac{4 k^2 m_{\chi}^2}{m_a^2 + m_{\chi}^2 + 2 m_{\chi} \omega} + m_a^2 - 2 m_{\chi} \omega \right)^3 \right) - \frac{m_a^4 m_{\chi}^2 (m_a^2 - 2 m_{\chi} \omega) (m_a^2 + 2 m_{\chi} \omega)^2}{\frac{4 k^2 m_{\chi}^2}{m_a^2 + m_{\chi}^2 + 2 m_{\chi} \omega} + m_a^2 - 2 m_{\chi} \omega} \notag \\
&\quad - 2 m_{\chi}^2 \left(m_a^8 - 2 m_a^4 m_{\chi}^2 \omega^2 + 8 m_a^2 m_{\chi}^3 \omega^3 + 8 m_{\chi}^4 \omega^4 \right) \notag \\
&\quad \times \log\left(-\frac{-2 m_{\chi}^2 (\omega (m_{\chi} + 2 \omega) - 2 k^2) + m_a^4 + m_a^2 m_{\chi}^2}{M_a^2 + m_{\chi} (m_{\chi} + 2 \omega)}\right) \notag \\
&\quad - ( -m_a^6 + m_a^4 m_{\chi} (m_{\chi} - 2 \omega) + 4 m_a^2 m_{\chi}^2 \omega (m_{\chi} + \omega) \notag \\
&\quad + 2 m_{\chi}^3 \omega^2 (3 m_{\chi} + 4 \omega)) \left(\frac{4 k^2 m_{\chi}^2}{m_a^2 + m_{\chi}^2 + 2 m_{\chi} \omega} + m_a^2 - 2 m_{\chi} \omega \right)^2 \notag \\
&\quad - (m_a^8 - 4 m_a^6 m_{\chi}^2 - 8 m_a^4 m_{\chi}^2 \omega^2 + 20 m_a^2 m_{\chi}^4 \omega^2 \notag \\
&\quad + 8 m_{\chi}^4 \omega^3 (3 m_{\chi} + 2 \omega)) \left(\frac{4 k^2 m_{\chi}^2}{m_a^2 + m_{\chi}^2 + 2 m_{\chi} \omega} + m_a^2 - 2 m_{\chi} \omega \right) \notag \\
&\quad + \frac{1}{3} \left( m_a^{10} - m_a^8 m_{\chi} (5 m_{\chi} + 2 \omega) + 8 m_a^6 m_{\chi}^2 \omega (4 m_{\chi} - \omega) \right. \notag \\
&\quad + 2 m_a^4 m_{\chi}^3 \omega^2 (27 m_{\chi} + 8 \omega) - 8 m_a^2 m_{\chi}^4 \omega^3 (7 m_{\chi} - 2 \omega) \notag \\
&\quad + 6 m_{\chi}^2 \left( m_a^8 - 2 m_a^4 m_{\chi}^2 \omega^2 + 8 m_a^2 m_{\chi}^3 \omega^3 + 8 m_{\chi}^4 \omega^4 \right) \notag \\
&\quad \left. \times \log(2 m_{\chi} \omega - m_a^2) - 8 m_{\chi}^5 \omega^4 (11 m_{\chi} + 4 \omega) \right)
\Bigg),
\end{eqnarray}
where $k$ and $\omega$ respectively represent the four-momentum and the energy of the thermal bath particles involved in scattering with DM. 

\end{widetext}
\bibliography{biblio}

@article{Planck:2018vyg,
    author = "Aghanim, N. and others",
    collaboration = "Planck",
    title = "{Planck 2018 results. VI. Cosmological parameters}",
    eprint = "1807.06209",
    archivePrefix = "arXiv",
    primaryClass = "astro-ph.CO",
    doi = "10.1051/0004-6361/201833910",
    journal = "Astron. Astrophys.",
    volume = "641",
    pages = "A6",
    year = "2020",
    note = "[Erratum: Astron.Astrophys. 652, C4 (2021)]"
}

@article{Belle:2017oht,
    author = "Grygier, J. and others",
    collaboration = "Belle",
    title = "{Search for $\boldsymbol{B\to h\nu\bar{\nu}}$ decays with semileptonic tagging at Belle}",
    eprint = "1702.03224",
    archivePrefix = "arXiv",
    primaryClass = "hep-ex",
    doi = "10.1103/PhysRevD.96.091101",
    journal = "Phys. Rev. D",
    volume = "96",
    number = "9",
    pages = "091101",
    year = "2017",
    note = "[Addendum: Phys.Rev.D 97, 099902 (2018)]"
}

@article{Abumusabh:2025zsr,
    author = "Abumusabh, Merna and Dujany, Giulio and Guadagnoli, Diego and Iohner, Axel and Toni, Claudio",
    title = "{The $B^+ \to K^+ ν\bar ν$ decay as a search for the QCD axion}",
    eprint = "2510.18953",
    archivePrefix = "arXiv",
    primaryClass = "hep-ph",
    reportNumber = "LAPTH-036/25",
    month = "10",
    year = "2025"
}

@article{Aliev:2025hyp,
    author = "Aliev, T. M. and Elpe, A. and Turan, I. and Selbuz, L.",
    title = "{Explaining Belle Data on $B\to K^{(*)}\nu\bar{\nu}$ Decays via Dark Photon Resonances}",
    eprint = "2503.22347",
    archivePrefix = "arXiv",
    primaryClass = "hep-ph",
    month = "3",
    year = "2025"
}

@article{Hochberg:2018rjs,
    author = "Hochberg, Yonit and Kuflik, Eric and Mcgehee, Robert and Murayama, Hitoshi and Schutz, Katelin",
    title = "{Strongly interacting massive particles through the axion portal}",
    eprint = "1806.10139",
    archivePrefix = "arXiv",
    primaryClass = "hep-ph",
    reportNumber = "DESY-18-101, IPMU18-0114",
    doi = "10.1103/PhysRevD.98.115031",
    journal = "Phys. Rev. D",
    volume = "98",
    number = "11",
    pages = "115031",
    year = "2018"
}

@article{Chen:2025npb,
    author = "Chen, Chuan-Hung and Chiang, Cheng-Wei and de la Vega, Leon M. G.",
    title = "{Leptoquark-mediated Dirac neutrino mass and its impact on $B \to K \nu \bar{\nu}$ and $K \to \pi \nu \bar{\nu}$ decays}",
    eprint = "2503.22431",
    archivePrefix = "arXiv",
    primaryClass = "hep-ph",
    month = "3",
    year = "2025"
}

@article{Bolton:2025fsq,
    author = "Bolton, Patrick D. and Fajfer, Svjetlana and Kamenik, Jernej F. and Novoa-Brunet, Mart\'\i{}n",
    title = "{Impact of new invisible particles on $B\to K^{(*)} E_{\rm miss}$ observables}",
    eprint = "2503.19025",
    archivePrefix = "arXiv",
    primaryClass = "hep-ph",
    month = "3",
    year = "2025"
}

@article{NA62:2021zjw,
    author = "Cortina Gil, Eduardo and others",
    collaboration = "NA62",
    title = "{Measurement of the very rare K$^{+}$\textrightarrow{}$ {\pi}^{+}\nu \overline{\nu} $ decay}",
    eprint = "2103.15389",
    archivePrefix = "arXiv",
    primaryClass = "hep-ex",
    doi = "10.1007/JHEP06(2021)093",
    journal = "JHEP",
    volume = "06",
    pages = "093",
    year = "2021"
}

@article{FASER:2024bbl,
    author = "Mammen Abraham, Roshan and others",
    collaboration = "FASER",
    title = "{Shining light on the dark sector: search for axion-like particles and other new physics in photonic final states with FASER}",
    eprint = "2410.10363",
    archivePrefix = "arXiv",
    primaryClass = "hep-ex",
    reportNumber = "CERN-EP-2024-262",
    doi = "10.1007/JHEP01(2025)199",
    journal = "JHEP",
    volume = "01",
    pages = "199",
    year = "2025"
}

@article{Georgi:1986df,
    author = "Georgi, Howard and Kaplan, David B. and Randall, Lisa",
    title = "{Manifesting the Invisible Axion at Low-energies}",
    reportNumber = "HUTP-86/A004",
    doi = "10.1016/0370-2693(86)90688-X",
    journal = "Phys. Lett. B",
    volume = "169",
    pages = "73--78",
    year = "1986"
}

@article{Brivio:2017ije,
    author = "Brivio, I. and Gavela, M. B. and Merlo, L. and Mimasu, K. and No, J. M. and del Rey, R. and Sanz, V.",
    title = "{ALPs Effective Field Theory and Collider Signatures}",
    eprint = "1701.05379",
    archivePrefix = "arXiv",
    primaryClass = "hep-ph",
    reportNumber = "IFT-UAM-CSIC-16-141, KCL-PH-TH-2016-72, FTUAM-16-49, CP3-17-04",
    doi = "10.1140/epjc/s10052-017-5111-3",
    journal = "Eur. Phys. J. C",
    volume = "77",
    number = "8",
    pages = "572",
    year = "2017"
}

@article{Ren:2021prq,
    author = "Ren, Jie and Wang, Daohan and Wu, Lei and Yang, Jin Min and Zhang, Mengchao",
    title = "{Detecting an axion-like particle with machine learning at the LHC}",
    eprint = "2106.07018",
    archivePrefix = "arXiv",
    primaryClass = "hep-ph",
    doi = "10.1007/JHEP11(2021)138",
    journal = "JHEP",
    volume = "11",
    pages = "138",
    year = "2021"
}

@article{Lu:2023ryd,
    author = "Lu, Chih-Ting and Luo, Xiaoyi and Wei, Xinqi",
    title = "{Exploring muonphilic ALPs at muon colliders}",
    eprint = "2303.03110",
    archivePrefix = "arXiv",
    primaryClass = "hep-ph",
    doi = "10.1088/1674-1137/ace424",
    journal = "Chin. Phys. C",
    volume = "47",
    number = "10",
    pages = "103102",
    year = "2023"
}

@article{Cheung:2023nzg,
    author = "Cheung, Kingman and Ouseph, C. J.",
    title = "{Axionlike particle search at Higgs factories}",
    eprint = "2303.16514",
    archivePrefix = "arXiv",
    primaryClass = "hep-ph",
    doi = "10.1103/PhysRevD.108.035003",
    journal = "Phys. Rev. D",
    volume = "108",
    number = "3",
    pages = "035003",
    year = "2023"
}

@article{Izaguirre:2016dfi,
    author = "Izaguirre, Eder and Lin, Tongyan and Shuve, Brian",
    title = "{Searching for Axionlike Particles in Flavor-Changing Neutral Current Processes}",
    eprint = "1611.09355",
    archivePrefix = "arXiv",
    primaryClass = "hep-ph",
    reportNumber = "SLAC-PUB-16876",
    doi = "10.1103/PhysRevLett.118.111802",
    journal = "Phys. Rev. Lett.",
    volume = "118",
    number = "11",
    pages = "111802",
    year = "2017"
}

@article{Ferber:2022rsf,
    author = {Ferber, Torben and Filimonova, Anastasiia and Sch\"afer, Ruth and Westhoff, Susanne},
    title = "{Displaced or invisible? ALPs from B decays at Belle II}",
    eprint = "2201.06580",
    archivePrefix = "arXiv",
    primaryClass = "hep-ph",
    reportNumber = "P3H-22-005, Nikhef 2022-001",
    doi = "10.1007/JHEP04(2023)131",
    journal = "JHEP",
    volume = "04",
    pages = "131",
    year = "2023"
}

@article{Liu:2023kat,
    author = "Liu, Yu and Liu, Xuewen and Zhu, Bin",
    title = "{Early kinetic decoupling effect on the forbidden dark matter annihilations into standard model particles}",
    eprint = "2301.12199",
    archivePrefix = "arXiv",
    primaryClass = "hep-ph",
    doi = "10.1103/PhysRevD.107.115009",
    journal = "Phys. Rev. D",
    volume = "107",
    number = "11",
    pages = "115009",
    year = "2023"
}

@article{Zhu:2021vlz,
    author = "Zhu, Bin and Liu, Xuewen",
    title = "{Probing the flavor-specific scalar mediator for the muon (g-2) deviation, the proton radius puzzle and the light dark matter production}",
    eprint = "2104.03238",
    archivePrefix = "arXiv",
    primaryClass = "hep-ph",
    doi = "10.1007/s11433-021-1819-5",
    journal = "Sci. China Phys. Mech. Astron.",
    volume = "65",
    number = "3",
    pages = "231011",
    year = "2022"
}

@article{Duerr:2020muu,
    author = "Duerr, Michael and Ferber, Torben and Garcia-Cely, Camilo and Hearty, Christopher and Schmidt-Hoberg, Kai",
    title = "{Long-lived Dark Higgs and Inelastic Dark Matter at Belle II}",
    eprint = "2012.08595",
    archivePrefix = "arXiv",
    primaryClass = "hep-ph",
    reportNumber = "DESY-20-226",
    doi = "10.1007/JHEP04(2021)146",
    journal = "JHEP",
    volume = "04",
    pages = "146",
    year = "2021"
}

@article{Jaeckel:2015jla,
    author = "Jaeckel, Joerg and Spannowsky, Michael",
    title = "{Probing MeV to 90 GeV axion-like particles with LEP and LHC}",
    eprint = "1509.00476",
    archivePrefix = "arXiv",
    primaryClass = "hep-ph",
    doi = "10.1016/j.physletb.2015.12.037",
    journal = "Phys. Lett. B",
    volume = "753",
    pages = "482--487",
    year = "2016"
}

@article{Belle-II:2018jsg,
    author = "Altmannshofer, W. and others",
    editor = "Kou, E. and Urquijo, P.",
    collaboration = "Belle-II",
    title = "{The Belle II Physics Book}",
    eprint = "1808.10567",
    archivePrefix = "arXiv",
    primaryClass = "hep-ex",
    reportNumber = "KEK Preprint 2018-27, BELLE2-PUB-PH-2018-001, FERMILAB-PUB-18-398-T, JLAB-THY-18-2780, INT-PUB-18-047, UWThPh 2018-26",
    doi = "10.1093/ptep/ptz106",
    journal = "PTEP",
    volume = "2019",
    number = "12",
    pages = "123C01",
    year = "2019",
    note = "[Erratum: PTEP 2020, 029201 (2020)]"
}

@article{vandenAarssen:2012ag,
    author = "van den Aarssen, Laura G. and Bringmann, Torsten and Goedecke, Yasar C",
    title = "{Thermal decoupling and the smallest subhalo mass in dark matter models with Sommerfeld-enhanced annihilation rates}",
    eprint = "1202.5456",
    archivePrefix = "arXiv",
    primaryClass = "hep-ph",
    doi = "10.1103/PhysRevD.85.123512",
    journal = "Phys. Rev. D",
    volume = "85",
    pages = "123512",
    year = "2012"
}

@article{Binder:2021bmg,
    author = "Binder, Tobias and Bringmann, Torsten and Gustafsson, Michael and Hryczuk, Andrzej",
    title = "{Dark matter relic abundance beyond kinetic equilibrium}",
    eprint = "2103.01944",
    archivePrefix = "arXiv",
    primaryClass = "hep-ph",
    doi = "10.1140/epjc/s10052-021-09357-5",
    journal = "Eur. Phys. J. C",
    volume = "81",
    pages = "577",
    year = "2021"
}

@article{Duan:2024urq,
    author = "Duan, Xin-Chen and Ramos, Raymundo and Tsai, Yue-Lin Sming",
    title = "{Relic density and temperature evolution of a light dark sector}",
    eprint = "2404.12019",
    archivePrefix = "arXiv",
    primaryClass = "hep-ph",
    doi = "10.1103/PhysRevD.110.063535",
    journal = "Phys. Rev. D",
    volume = "110",
    number = "6",
    pages = "063535",
    year = "2024"
}

@article{Aboubrahim:2023yag,
    author = "Aboubrahim, Amin and Klasen, Michael and Wiggering, Luca Paolo",
    title = "{Forbidden dark matter annihilation into leptons with full collision terms}",
    eprint = "2306.07753",
    archivePrefix = "arXiv",
    primaryClass = "hep-ph",
    reportNumber = "MS-TP-23-15",
    doi = "10.1088/1475-7516/2023/08/075",
    journal = "JCAP",
    volume = "08",
    pages = "075",
    year = "2023"
}

@article{Dolan:2017osp,
    author = "Dolan, Matthew J. and Ferber, Torben and Hearty, Christopher and Kahlhoefer, Felix and Schmidt-Hoberg, Kai",
    title = "{Revised constraints and Belle II sensitivity for visible and invisible axion-like particles}",
    eprint = "1709.00009",
    archivePrefix = "arXiv",
    primaryClass = "hep-ph",
    reportNumber = "DESY-17-127",
    doi = "10.1007/JHEP12(2017)094",
    journal = "JHEP",
    volume = "12",
    pages = "094",
    year = "2017",
    note = "[Erratum: JHEP 03, 190 (2021)]"
}

@article{BaBar:2017tiz,
    author = "Lees, J. P. and others",
    collaboration = "BaBar",
    title = "{Search for Invisible Decays of a Dark Photon Produced in ${e}^{+}{e}^{-}$ Collisions at BaBar}",
    eprint = "1702.03327",
    archivePrefix = "arXiv",
    primaryClass = "hep-ex",
    reportNumber = "BABAR-PUB-17-001, SLAC-PUB-16923",
    doi = "10.1103/PhysRevLett.119.131804",
    journal = "Phys. Rev. Lett.",
    volume = "119",
    number = "13",
    pages = "131804",
    year = "2017"
}

@article{Gondolo:1990dk,
    author = "Gondolo, Paolo and Gelmini, Graciela",
    title = "{Cosmic abundances of stable particles: Improved analysis}",
    reportNumber = "UCLA-90-TEP-68",
    doi = "10.1016/0550-3213(91)90438-4",
    journal = "Nucl. Phys. B",
    volume = "360",
    pages = "145--179",
    year = "1991"
}

@article{Binder:2017rgn,
    author = "Binder, Tobias and Bringmann, Torsten and Gustafsson, Michael and Hryczuk, Andrzej",
    title = "{Early kinetic decoupling of dark matter: when the standard way of calculating the thermal relic density fails}",
    eprint = "1706.07433",
    archivePrefix = "arXiv",
    primaryClass = "astro-ph.CO",
    doi = "10.1103/PhysRevD.96.115010",
    journal = "Phys. Rev. D",
    volume = "96",
    number = "11",
    pages = "115010",
    year = "2017",
    note = "[Erratum: Phys.Rev.D 101, 099901 (2020)]"
}

@article{Binder:2016pnr,
    author = "Binder, Tobias and Covi, Laura and Kamada, Ayuki and Murayama, Hitoshi and Takahashi, Tomo and Yoshida, Naoki",
    title = "{Matter Power Spectrum in Hidden Neutrino Interacting Dark Matter Models: A Closer Look at the Collision Term}",
    eprint = "1602.07624",
    archivePrefix = "arXiv",
    primaryClass = "hep-ph",
    doi = "10.1088/1475-7516/2016/11/043",
    journal = "JCAP",
    volume = "11",
    pages = "043",
    year = "2016"
}

@article{Bertschinger:2006nq,
    author = "Bertschinger, Edmund",
    title = "{The Effects of Cold Dark Matter Decoupling and Pair Annihilation on Cosmological Perturbations}",
    eprint = "astro-ph/0607319",
    archivePrefix = "arXiv",
    doi = "10.1103/PhysRevD.74.063509",
    journal = "Phys. Rev. D",
    volume = "74",
    pages = "063509",
    year = "2006"
}

@article{Bringmann:2006mu,
    author = "Bringmann, Torsten and Hofmann, Stefan",
    title = "{Thermal decoupling of WIMPs from first principles}",
    eprint = "hep-ph/0612238",
    archivePrefix = "arXiv",
    doi = "10.1088/1475-7516/2007/04/016",
    journal = "JCAP",
    volume = "04",
    pages = "016",
    year = "2007",
    note = "[Erratum: JCAP 03, E02 (2016)]"
}

@article{Bringmann:2009vf,
    author = "Bringmann, Torsten",
    title = "{Particle Models and the Small-Scale Structure of Dark Matter}",
    eprint = "0903.0189",
    archivePrefix = "arXiv",
    primaryClass = "astro-ph.CO",
    doi = "10.1088/1367-2630/11/10/105027",
    journal = "New J. Phys.",
    volume = "11",
    pages = "105027",
    year = "2009"
}

@article{Gondolo:2012vh,
    author = "Gondolo, Paolo and Hisano, Junji and Kadota, Kenji",
    title = "{The Effect of quark interactions on dark matter kinetic decoupling and the mass of the smallest dark halos}",
    eprint = "1205.1914",
    archivePrefix = "arXiv",
    primaryClass = "hep-ph",
    doi = "10.1103/PhysRevD.86.083523",
    journal = "Phys. Rev. D",
    volume = "86",
    pages = "083523",
    year = "2012"
}

@article{Brummer:2019inq,
    author = {Br\"ummer, F.},
    title = "{Coscattering in next-to-minimal dark matter and split supersymmetry}",
    eprint = "1910.01549",
    archivePrefix = "arXiv",
    primaryClass = "hep-ph",
    doi = "10.1007/JHEP01(2020)113",
    journal = "JHEP",
    volume = "01",
    pages = "113",
    year = "2020"
}

@article{Ala-Mattinen:2019mpa,
    author = "Ala-Mattinen, Kalle and Kainulainen, Kimmo",
    title = "{Precision calculations of dark matter relic abundance}",
    eprint = "1912.02870",
    archivePrefix = "arXiv",
    primaryClass = "hep-ph",
    reportNumber = "CERN-TH-2019-214",
    doi = "10.1088/1475-7516/2020/09/040",
    journal = "JCAP",
    volume = "09",
    pages = "040",
    year = "2020"
}

@article{Abe:2020obo,
    author = "Abe, Tomohiro",
    title = "{Effect of the early kinetic decoupling in a fermionic dark matter model}",
    eprint = "2004.10041",
    archivePrefix = "arXiv",
    primaryClass = "hep-ph",
    doi = "10.1103/PhysRevD.102.035018",
    journal = "Phys. Rev. D",
    volume = "102",
    number = "3",
    pages = "035018",
    year = "2020"
}

@article{Hryczuk:2022gay,
    author = "Hryczuk, Andrzej and Laletin, Maxim",
    title = "{Impact of dark matter self-scattering on its relic abundance}",
    eprint = "2204.07078",
    archivePrefix = "arXiv",
    primaryClass = "hep-ph",
    doi = "10.1103/PhysRevD.106.023007",
    journal = "Phys. Rev. D",
    volume = "106",
    number = "2",
    pages = "023007",
    year = "2022"
}

@article{Hryczuk:2021qtz,
    author = "Hryczuk, Andrzej and Laletin, Maxim",
    title = "{Dark matter freeze-in from semi-production}",
    eprint = "2104.05684",
    archivePrefix = "arXiv",
    primaryClass = "hep-ph",
    doi = "10.1007/JHEP06(2021)026",
    journal = "JHEP",
    volume = "06",
    pages = "026",
    year = "2021"
}

@article{Abe:2021jcz,
    author = "Abe, Tomohiro",
    title = "{Early kinetic decoupling and a pseudo-Nambu-Goldstone dark matter model}",
    eprint = "2106.01956",
    archivePrefix = "arXiv",
    primaryClass = "hep-ph",
    doi = "10.1103/PhysRevD.104.035025",
    journal = "Phys. Rev. D",
    volume = "104",
    number = "3",
    pages = "035025",
    year = "2021"
}

@article{Du:2021jcj,
    author = "Du, Yong and Huang, Fei and Li, Hao-Lin and Li, Yuan-Zhen and Yu, Jiang-Hao",
    title = "{Revisiting dark matter freeze-in and freeze-out through phase-space distribution}",
    eprint = "2111.01267",
    archivePrefix = "arXiv",
    primaryClass = "hep-ph",
    reportNumber = "UCI-HEP-TR-2021-27, Report-no: UCI-HEP-TR-2021-27",
    doi = "10.1088/1475-7516/2022/04/012",
    journal = "JCAP",
    volume = "04",
    number = "04",
    pages = "012",
    year = "2022"
}

@article{Ala-Mattinen:2022nuj,
    author = "Ala-Mattinen, Kalle and Heikinheimo, Matti and Kainulainen, Kimmo and Tuominen, Kimmo",
    title = "{Momentum distributions of cosmic relics: Improved analysis}",
    eprint = "2201.06456",
    archivePrefix = "arXiv",
    primaryClass = "hep-ph",
    reportNumber = "HIP-2021-37/TH",
    doi = "10.1103/PhysRevD.105.123005",
    journal = "Phys. Rev. D",
    volume = "105",
    number = "12",
    pages = "123005",
    year = "2022"
}

@article{Belle-II:2023esi,
    author = "Adachi, I. and others",
    collaboration = "Belle-II",
    title = "{Evidence for B+\textrightarrow{}K+\ensuremath{\nu}\ensuremath{\nu}\textasciimacron{} decays}",
    eprint = "2311.14647",
    archivePrefix = "arXiv",
    primaryClass = "hep-ex",
    reportNumber = "Belle II Preprint 2023-017, KEK Preprint 2023-35",
    doi = "10.1103/PhysRevD.109.112006",
    journal = "Phys. Rev. D",
    volume = "109",
    number = "11",
    pages = "112006",
    year = "2024"
}

@article{Athron:2023hmz,
    author = "Athron, Peter and Martinez, R. and Sierra, Cristian",
    title = "{B meson anomalies and large $ {B}^{+}\to {K}^{+}\nu \overline{\nu} $ in non-universal U(1)$^{′}$ models}",
    eprint = "2308.13426",
    archivePrefix = "arXiv",
    primaryClass = "hep-ph",
    doi = "10.1007/JHEP02(2024)121",
    journal = "JHEP",
    volume = "02",
    pages = "121",
    year = "2024"
}

@article{Bause:2023mfe,
    author = "Bause, Rigo and Gisbert, Hector and Hiller, Gudrun",
    title = "{Implications of an enhanced B\textrightarrow{}K\ensuremath{\nu}\ensuremath{\nu}\textasciimacron{} branching ratio}",
    eprint = "2309.00075",
    archivePrefix = "arXiv",
    primaryClass = "hep-ph",
    doi = "10.1103/PhysRevD.109.015006",
    journal = "Phys. Rev. D",
    volume = "109",
    number = "1",
    pages = "015006",
    year = "2024"
}

@article{Allwicher:2023xba,
    author = "Allwicher, Lukas and Becirevic, Damir and Piazza, Gioacchino and Rosauro-Alcaraz, Salvador and Sumensari, Olcyr",
    title = "{Understanding the first measurement of B(B\textrightarrow{}K\ensuremath{\nu}\ensuremath{\nu}\textasciimacron{})}",
    eprint = "2309.02246",
    archivePrefix = "arXiv",
    primaryClass = "hep-ph",
    doi = "10.1016/j.physletb.2023.138411",
    journal = "Phys. Lett. B",
    volume = "848",
    pages = "138411",
    year = "2024"
}

@article{Abdughani:2023dlr,
    author = "Abdughani, Murat and Reyimuaji, Yakefu",
    title = "{Constraining light dark matter and mediator with B+\textrightarrow{}K+\ensuremath{\nu}\ensuremath{\nu}\textasciimacron{} data}",
    eprint = "2309.03706",
    archivePrefix = "arXiv",
    primaryClass = "hep-ph",
    doi = "10.1103/PhysRevD.110.055013",
    journal = "Phys. Rev. D",
    volume = "110",
    number = "5",
    pages = "055013",
    year = "2024"
}

@article{He:2023bnk,
    author = "He, Xiao-Gang and Ma, Xiao-Dong and Valencia, German",
    title = "{Revisiting models that enhance B+\textrightarrow{}K+\ensuremath{\nu}\ensuremath{\nu}\textasciimacron{} in light of the new Belle II measurement}",
    eprint = "2309.12741",
    archivePrefix = "arXiv",
    primaryClass = "hep-ph",
    doi = "10.1103/PhysRevD.109.075019",
    journal = "Phys. Rev. D",
    volume = "109",
    number = "7",
    pages = "075019",
    year = "2024"
}

@article{Datta:2023iln,
    author = "Datta, Alakabha and Marfatia, Danny and Mukherjee, Lopamudra",
    title = "{B\textrightarrow{}K\ensuremath{\nu}\ensuremath{\nu}\textasciimacron{}, MiniBooNE and muon g-2 anomalies from a dark sector}",
    eprint = "2310.15136",
    archivePrefix = "arXiv",
    primaryClass = "hep-ph",
    doi = "10.1103/PhysRevD.109.L031701",
    journal = "Phys. Rev. D",
    volume = "109",
    number = "3",
    pages = "L031701",
    year = "2024"
}

@article{Altmannshofer:2023hkn,
    author = "Altmannshofer, Wolfgang and Crivellin, Andreas and Haigh, Huw and Inguglia, Gianluca and Martin Camalich, Jorge",
    title = "{Light new physics in B\textrightarrow{}K(*)\ensuremath{\nu}\ensuremath{\nu}\textasciimacron{}?}",
    eprint = "2311.14629",
    archivePrefix = "arXiv",
    primaryClass = "hep-ph",
    reportNumber = "PSI-PR-23-46, ZU-TH 77/23",
    doi = "10.1103/PhysRevD.109.075008",
    journal = "Phys. Rev. D",
    volume = "109",
    number = "7",
    pages = "075008",
    year = "2024"
}

@article{McKeen:2023uzo,
    author = "McKeen, David and Ng, John N. and Tuckler, Douglas",
    title = "{Higgs portal interpretation of the Belle II B+\textrightarrow{}K+\ensuremath{\nu}\ensuremath{\nu} measurement}",
    eprint = "2312.00982",
    archivePrefix = "arXiv",
    primaryClass = "hep-ph",
    doi = "10.1103/PhysRevD.109.075006",
    journal = "Phys. Rev. D",
    volume = "109",
    number = "7",
    pages = "075006",
    year = "2024"
}

@article{Fridell:2023ssf,
    author = "Fridell, K\r{a}re and Ghosh, Mitrajyoti and Okui, Takemichi and Tobioka, Kohsaku",
    title = "{Decoding the B\textrightarrow{}K\ensuremath{\nu}\ensuremath{\nu} excess at Belle II: Kinematics, operators, and masses}",
    eprint = "2312.12507",
    archivePrefix = "arXiv",
    primaryClass = "hep-ph",
    reportNumber = "KEK-TH-2587",
    doi = "10.1103/PhysRevD.109.115006",
    journal = "Phys. Rev. D",
    volume = "109",
    number = "11",
    pages = "115006",
    year = "2024"
}

@article{Ho:2024cwk,
    author = "Ho, Shu-Yu and Kim, Jongkuk and Ko, Pyungwon",
    title = "{Recent $B^+ \!\to K^+\nu\bar{\nu}$ Excess and Muon $g-2$ Illuminating Light Dark Sector with Higgs Portal}",
    eprint = "2401.10112",
    archivePrefix = "arXiv",
    primaryClass = "hep-ph",
    reportNumber = "KIAS-24003",
    month = "1",
    year = "2024"
}

@article{Chen:2024jlj,
    author = "Chen, Feng-Zhi and Wen, Qiaoyi and Xu, Fanrong",
    title = "{Correlating $B\rightarrow K^{(*)} \nu \bar{\nu }$ and flavor anomalies in SMEFT}",
    eprint = "2401.11552",
    archivePrefix = "arXiv",
    primaryClass = "hep-ph",
    doi = "10.1140/epjc/s10052-024-13425-x",
    journal = "Eur. Phys. J. C",
    volume = "84",
    number = "10",
    pages = "1012",
    year = "2024"
}

@article{Gabrielli:2024wys,
    author = {Gabrielli, Emidio and Marzola, Luca and M\"u\"ursepp, Kristjan and Raidal, Martti},
    title = "{Explaining the $B^+\rightarrow K^+ \nu \bar{\nu }$ excess via a massless dark photon}",
    eprint = "2402.05901",
    archivePrefix = "arXiv",
    primaryClass = "hep-ph",
    doi = "10.1140/epjc/s10052-024-12818-2",
    journal = "Eur. Phys. J. C",
    volume = "84",
    number = "5",
    pages = "460",
    year = "2024"
}

@article{Hou:2024vyw,
    author = "Hou, Biao-Feng and Li, Xin-Qiang and Shen, Meng and Yang, Ya-Dong and Yuan, Xing-Bo",
    title = "{Deciphering the Belle II data on $ B\to K\nu \overline{\nu} $ decay in the (dark) SMEFT with minimal flavour violation}",
    eprint = "2402.19208",
    archivePrefix = "arXiv",
    primaryClass = "hep-ph",
    doi = "10.1007/JHEP06(2024)172",
    journal = "JHEP",
    volume = "06",
    pages = "172",
    year = "2024"
}

@article{Chen:2024cll,
    author = "Chen, Chuan-Hung and Chiang, Cheng-Wei",
    title = "{Rare B and K decays in a scotogenic model}",
    eprint = "2403.02897",
    archivePrefix = "arXiv",
    primaryClass = "hep-ph",
    doi = "10.1103/PhysRevD.110.075036",
    journal = "Phys. Rev. D",
    volume = "110",
    number = "7",
    pages = "075036",
    year = "2024"
}

@article{He:2024iju,
    author = "He, Xiao-Gang and Ma, Xiao-Dong and Schmidt, Michael A. and Valencia, German and Volkas, Raymond R.",
    title = "{Scalar dark matter explanation of the excess in the Belle II B$^{+}$\textrightarrow{} K$^{+}$+ invisible measurement}",
    eprint = "2403.12485",
    archivePrefix = "arXiv",
    primaryClass = "hep-ph",
    reportNumber = "CPPC-2024-04",
    doi = "10.1007/JHEP07(2024)168",
    journal = "JHEP",
    volume = "07",
    pages = "168",
    year = "2024"
}

@article{Bolton:2024egx,
    author = "Bolton, Patrick D. and Fajfer, Svjetlana and Kamenik, Jernej F. and Novoa-Brunet, Mart\'\i{}n",
    title = "{Signatures of light new particles in B\textrightarrow{}K(*)Emiss}",
    eprint = "2403.13887",
    archivePrefix = "arXiv",
    primaryClass = "hep-ph",
    doi = "10.1103/PhysRevD.110.055001",
    journal = "Phys. Rev. D",
    volume = "110",
    number = "5",
    pages = "055001",
    year = "2024"
}

@article{Marzocca:2024hua,
    author = "Marzocca, David and Nardecchia, Marco and Stanzione, Alfredo and Toni, Claudio",
    title = "{Implications of $B \rightarrow K \nu {\bar{\nu }}$ under rank-one flavor violation hypothesis}",
    eprint = "2404.06533",
    archivePrefix = "arXiv",
    primaryClass = "hep-ph",
    doi = "10.1140/epjc/s10052-024-13534-7",
    journal = "Eur. Phys. J. C",
    volume = "84",
    number = "11",
    pages = "1217",
    year = "2024"
}

@article{Rosauro-Alcaraz:2024mvx,
    author = "Rosauro-Alcaraz, S. and Leal, L. P. S.",
    title = "{Disentangling left and right-handed neutrino effects in $B\rightarrow K^{(*)}\nu \nu $}",
    eprint = "2404.17440",
    archivePrefix = "arXiv",
    primaryClass = "hep-ph",
    doi = "10.1140/epjc/s10052-024-13104-x",
    journal = "Eur. Phys. J. C",
    volume = "84",
    number = "8",
    pages = "795",
    year = "2024"
}

@article{Kim:2024tsm,
    author = "Kim, C. S. and Sahoo, Dibyakrupa and Vishnudath, K. N.",
    title = "{Searching for signatures of new physics in $\varvec{B \rightarrow K \, \nu \, \overline{\nu }}$ to distinguish between Dirac and Majorana neutrinos}",
    eprint = "2405.17341",
    archivePrefix = "arXiv",
    primaryClass = "hep-ph",
    doi = "10.1140/epjc/s10052-024-13262-y",
    journal = "Eur. Phys. J. C",
    volume = "84",
    number = "9",
    pages = "882",
    year = "2024"
}

@article{Hati:2024ppg,
    author = "Hati, Chandan and Leite, Julio and Nath, Newton and Valle, Jos\'e W. F.",
    title = "{The QCD axion, colour-mediated neutrino masses, and $B^+\to K^+ + E_{\text{miss}}$ anomaly}",
    eprint = "2408.00060",
    archivePrefix = "arXiv",
    primaryClass = "hep-ph",
    month = "7",
    year = "2024"
}

@article{Allwicher:2024ncl,
    author = "Allwicher, Lukas and Bordone, Marzia and Isidori, Gino and Piazza, Gioacchino and Stanzione, Alfredo",
    title = "{Probing third-generation New Physics with $K\to \pi \nu\bar\nu$ and $B\to K^{(*)} \nu\bar\nu$}",
    eprint = "2410.21444",
    archivePrefix = "arXiv",
    primaryClass = "hep-ph",
    month = "10",
    year = "2024"
}

@article{Becirevic:2024iyi,
    author = "Be\v{c}irevi\'c, Damir and Fajfer, Svjetlana and Ko\v{s}nik, Nejc and Pavi\v{c}i\'c, Lovre",
    title = "{Right-handed interactions in puzzling $B$-decays}",
    eprint = "2410.23257",
    archivePrefix = "arXiv",
    primaryClass = "hep-ph",
    month = "10",
    year = "2024"
}

@article{Altmannshofer:2024kxb,
    author = "Altmannshofer, Wolfgang and Roy, Shibasis",
    title = "{A joint explanation of the $B\to \pi K$ puzzle and the $B \to K \nu \bar{\nu}$ excess}",
    eprint = "2411.06592",
    archivePrefix = "arXiv",
    primaryClass = "hep-ph",
    month = "11",
    year = "2024"
}

@article{Buras:2024mnq,
    author = "Buras, Andrzej J. and Stangl, Peter",
    title = "{On the Interplay of Constraints from $B_s$, $D$, and $K$ Meson Mixing in $Z^\prime$ Models with Implications for $b\to s \nu\bar\nu$ Transitions}",
    eprint = "2412.14254",
    archivePrefix = "arXiv",
    primaryClass = "hep-ph",
    reportNumber = "AJB-24-3, CERN-TH-2024-218",
    month = "12",
    year = "2024"
}

@article{Hu:2024mgf,
    author = "Hu, Quan-Yi",
    title = "{Are the new particles heavy or light in $b \to s E_{\mathrm{miss}}$?}",
    eprint = "2412.19084",
    archivePrefix = "arXiv",
    primaryClass = "hep-ph",
    month = "12",
    year = "2024"
}

@article{Zhang:2024hkn,
    author = "Zhang, Chao-Qi and Sun, Jin and Xing, Zhi-Peng and Zhu, Rui-Lin",
    title = "{Probing $B^+ \to K^+$ semi-leptonic FCNC decay with new physics effects under PQCD approach}",
    eprint = "2501.00512",
    archivePrefix = "arXiv",
    primaryClass = "hep-ph",
    month = "12",
    year = "2024"
}

@article{Gola:2021abm,
    author = "Gola, Shivam and Mandal, Sanjoy and Sinha, Nita",
    title = "{ALP-portal majorana dark matter}",
    eprint = "2106.00547",
    archivePrefix = "arXiv",
    primaryClass = "hep-ph",
    doi = "10.1142/S0217751X22501317",
    journal = "Int. J. Mod. Phys. A",
    volume = "37",
    number = "22",
    pages = "2250131",
    year = "2022"
}

@article{Bharucha:2022lty,
    author = {Bharucha, A. and Br\"ummer, F. and Desai, N. and Mutzel, S.},
    title = "{Axion-like particles as mediators for dark matter: beyond freeze-out}",
    eprint = "2209.03932",
    archivePrefix = "arXiv",
    primaryClass = "hep-ph",
    doi = "10.1007/JHEP02(2023)141",
    journal = "JHEP",
    volume = "02",
    pages = "141",
    year = "2023"
}

@article{Ghosh:2023tyz,
    author = "Ghosh, Dilip Kumar and Ghoshal, Anish and Jeesun, Sk",
    title = "{Axion-like particle (ALP) portal freeze-in dark matter confronting ALP search experiments}",
    eprint = "2305.09188",
    archivePrefix = "arXiv",
    primaryClass = "hep-ph",
    doi = "10.1007/JHEP01(2024)026",
    journal = "JHEP",
    volume = "01",
    pages = "026",
    year = "2024"
}

@article{Graham:2015ouw,
    author = "Graham, Peter W. and Irastorza, Igor G. and Lamoreaux, Steven K. and Lindner, Axel and van Bibber, Karl A.",
    title = "{Experimental Searches for the Axion and Axion-Like Particles}",
    eprint = "1602.00039",
    archivePrefix = "arXiv",
    primaryClass = "hep-ex",
    doi = "10.1146/annurev-nucl-102014-022120",
    journal = "Ann. Rev. Nucl. Part. Sci.",
    volume = "65",
    pages = "485--514",
    year = "2015"
}

@article{Choi:2020rgn,
    author = "Choi, Kiwoon and Im, Sang Hui and Sub Shin, Chang",
    title = "{Recent Progress in the Physics of Axions and Axion-Like Particles}",
    eprint = "2012.05029",
    archivePrefix = "arXiv",
    primaryClass = "hep-ph",
    reportNumber = "CTPU-PTC-20-28",
    doi = "10.1146/annurev-nucl-120720-031147",
    journal = "Ann. Rev. Nucl. Part. Sci.",
    volume = "71",
    pages = "225--252",
    year = "2021"
}

@article{Peccei:1977hh,
    author = "Peccei, R. D. and Quinn, Helen R.",
    title = "{CP Conservation in the Presence of Instantons}",
    reportNumber = "ITP-568-STANFORD",
    doi = "10.1103/PhysRevLett.38.1440",
    journal = "Phys. Rev. Lett.",
    volume = "38",
    pages = "1440--1443",
    year = "1977"
}

@article{Weinberg:1977ma,
    author = "Weinberg, Steven",
    title = "{A New Light Boson?}",
    reportNumber = "HUTP-77/A074",
    doi = "10.1103/PhysRevLett.40.223",
    journal = "Phys. Rev. Lett.",
    volume = "40",
    pages = "223--226",
    year = "1978"
}

@article{Wilczek:1977pj,
    author = "Wilczek, Frank",
    title = "{Problem of Strong  $P$  and  $T$  Invariance in the Presence of Instantons}",
    reportNumber = "Print-77-0939 (COLUMBIA)",
    doi = "10.1103/PhysRevLett.40.279",
    journal = "Phys. Rev. Lett.",
    volume = "40",
    pages = "279--282",
    year = "1978"
}

@article{Kim:1979if,
    author = "Kim, Jihn E.",
    title = "{Weak Interaction Singlet and Strong CP Invariance}",
    reportNumber = "UPR-0120T",
    doi = "10.1103/PhysRevLett.43.103",
    journal = "Phys. Rev. Lett.",
    volume = "43",
    pages = "103",
    year = "1979"
}

@article{Preskill:1982cy,
    author = "Preskill, John and Wise, Mark B. and Wilczek, Frank",
    editor = "Srednicki, M. A.",
    title = "{Cosmology of the Invisible Axion}",
    reportNumber = "HUTP-82-A048, NSF-ITP-82-103",
    doi = "10.1016/0370-2693(83)90637-8",
    journal = "Phys. Lett. B",
    volume = "120",
    pages = "127--132",
    year = "1983"
}

@article{Abbott:1982af,
    author = "Abbott, L. F. and Sikivie, P.",
    editor = "Srednicki, M. A.",
    title = "{A Cosmological Bound on the Invisible Axion}",
    reportNumber = "PRINT-82-0695 (BRANDEIS)",
    doi = "10.1016/0370-2693(83)90638-X",
    journal = "Phys. Lett. B",
    volume = "120",
    pages = "133--136",
    year = "1983"
}

@article{Dine:1982ah,
    author = "Dine, Michael and Fischler, Willy",
    editor = "Srednicki, M. A.",
    title = "{The Not So Harmless Axion}",
    reportNumber = "UPR-0201T",
    doi = "10.1016/0370-2693(83)90639-1",
    journal = "Phys. Lett. B",
    volume = "120",
    pages = "137--141",
    year = "1983"
}

@article{Blumenhagen:2014gta,
    author = "Blumenhagen, Ralph and Plauschinn, Erik",
    title = "{Towards Universal Axion Inflation and Reheating in String Theory}",
    eprint = "1404.3542",
    archivePrefix = "arXiv",
    primaryClass = "hep-th",
    doi = "10.1016/j.physletb.2014.08.007",
    journal = "Phys. Lett. B",
    volume = "736",
    pages = "482--487",
    year = "2014"
}

@article{Kobayashi:2020ryx,
    author = "Kobayashi, Takeshi and Ubaldi, Lorenzo",
    title = "{Reheating-Induced Axion Dark Matter After Low Scale Inflation}",
    eprint = "2006.09389",
    archivePrefix = "arXiv",
    primaryClass = "hep-ph",
    doi = "10.1007/JHEP09(2020)052",
    journal = "JHEP",
    volume = "09",
    pages = "052",
    year = "2020"
}

@article{Guendelman:1991se,
    author = "Guendelman, E. I. and Owen, D. A.",
    title = "{Axion driven baryogenesis}",
    reportNumber = "WIS-91-53-PH",
    doi = "10.1016/0370-2693(92)90548-I",
    journal = "Phys. Lett. B",
    volume = "276",
    pages = "108--114",
    year = "1992"
}

@article{Domcke:2020kcp,
    author = "Domcke, Valerie and Ema, Yohei and Mukaida, Kyohei and Yamada, Masaki",
    title = "{Spontaneous Baryogenesis from Axions with Generic Couplings}",
    eprint = "2006.03148",
    archivePrefix = "arXiv",
    primaryClass = "hep-ph",
    reportNumber = "DESY 20-100, DESY-20-100, CERN-TH-2020-088, TU-1102",
    doi = "10.1007/JHEP08(2020)096",
    journal = "JHEP",
    volume = "08",
    pages = "096",
    year = "2020"
}

@article{Co:2024oek,
    author = "Co, Raymond T. and Fernandez, Nicolas and Ghalsasi, Akshay and Harigaya, Keisuke and Shelton, Jessie",
    title = "{Axion baryogenesis puts a new spin on the Hubble tension}",
    eprint = "2405.12268",
    archivePrefix = "arXiv",
    primaryClass = "hep-ph",
    reportNumber = "PITT-PACC-2402",
    doi = "10.1103/PhysRevD.110.083534",
    journal = "Phys. Rev. D",
    volume = "110",
    number = "8",
    pages = "083534",
    year = "2024"
}

@article{Nomura:2008ru,
    author = "Nomura, Yasunori and Thaler, Jesse",
    title = "{Dark Matter through the Axion Portal}",
    eprint = "0810.5397",
    archivePrefix = "arXiv",
    primaryClass = "hep-ph",
    doi = "10.1103/PhysRevD.79.075008",
    journal = "Phys. Rev. D",
    volume = "79",
    pages = "075008",
    year = "2009"
}

@article{Allen:2024ndv,
    author = "Allen, Stephanie and Blackburn, Albany and Cardenas, Oswaldo and Messenger, Zoe and Nguyen, Ngan H. and Shuve, Brian",
    title = "{Electroweak axion portal to dark matter}",
    eprint = "2405.02403",
    archivePrefix = "arXiv",
    primaryClass = "hep-ph",
    doi = "10.1103/PhysRevD.110.095010",
    journal = "Phys. Rev. D",
    volume = "110",
    number = "9",
    pages = "095010",
    year = "2024"
}

@article{Im:2019iwd,
    author = "Im, Sang Hui and Jeong, Kwang Sik",
    title = "{Freeze-in Axion-like Dark Matter}",
    eprint = "1907.07383",
    archivePrefix = "arXiv",
    primaryClass = "hep-ph",
    reportNumber = "PNUTP-19-A13",
    doi = "10.1016/j.physletb.2019.135044",
    journal = "Phys. Lett. B",
    volume = "799",
    pages = "135044",
    year = "2019"
}

@article{Arias:2025nub,
    author = "Arias, Paola and Diaz Saez, Bastian and Jaeckel, Joerg",
    title = "{Freezing-in the Pure Dark Axion Portal}",
    eprint = "2501.17234",
    archivePrefix = "arXiv",
    primaryClass = "hep-ph",
    month = "1",
    year = "2025"
}

@article{Guo:2009aj,
    author = "Guo, Wan-Lei and Wu, Yue-Liang",
    title = "{Enhancement of Dark Matter Annihilation via Breit-Wigner Resonance}",
    eprint = "0901.1450",
    archivePrefix = "arXiv",
    primaryClass = "hep-ph",
    doi = "10.1103/PhysRevD.79.055012",
    journal = "Phys. Rev. D",
    volume = "79",
    pages = "055012",
    year = "2009"
}

@article{Laine:2022ner,
    author = "Laine, M.",
    title = "{Resonant s-channel dark matter annihilation at NLO}",
    eprint = "2211.06008",
    archivePrefix = "arXiv",
    primaryClass = "hep-ph",
    doi = "10.1007/JHEP01(2023)157",
    journal = "JHEP",
    volume = "01",
    pages = "157",
    year = "2023"
}

@article{Biswas:2019lcp,
    author = "Biswas, Sanjoy and Chatterjee, Anirban and Gabrielli, Emidio and Mele, Barbara",
    title = "{Probing dark-axionlike particle portals at future $e^+e^-$ colliders}",
    eprint = "1906.10608",
    archivePrefix = "arXiv",
    primaryClass = "hep-ph",
    doi = "10.1103/PhysRevD.100.115040",
    journal = "Phys. Rev. D",
    volume = "100",
    number = "11",
    pages = "115040",
    year = "2019"
}

@article{Armando:2023zwz,
    author = "Armando, Giovanni and Panci, Paolo and Weiss, Joachim and Ziegler, Robert",
    title = "{Leptonic ALP portal to the dark sector}",
    eprint = "2310.05827",
    archivePrefix = "arXiv",
    primaryClass = "hep-ph",
    reportNumber = "TTP23-045, P3H-23-071, MPP-2023-240, FR-PHENO-2023-11",
    doi = "10.1103/PhysRevD.109.055029",
    journal = "Phys. Rev. D",
    volume = "109",
    number = "5",
    pages = "055029",
    year = "2024"
}

@article{Jodlowski:2024lab,
    author = "Jod\l{}owski, Krzysztof",
    title = "{Dark axion portal at $Z$ boson factories}",
    eprint = "2411.19196",
    archivePrefix = "arXiv",
    primaryClass = "hep-ph",
    month = "11",
    year = "2024"
}

@article{Lee:2012bq,
    author = "Lee, Hyun Min and Park, Myeonghun and Park, Wan-Il",
    title = "{Fermi Gamma Ray Line at 130 GeV from Axion-Mediated Dark Matter}",
    eprint = "1205.4675",
    archivePrefix = "arXiv",
    primaryClass = "hep-ph",
    reportNumber = "CERN-PH-TH-2012-136",
    doi = "10.1103/PhysRevD.86.103502",
    journal = "Phys. Rev. D",
    volume = "86",
    pages = "103502",
    year = "2012"
}

@article{Ibarra:2013eda,
    author = "Ibarra, Alejandro and Lee, Hyun Min and L\'opez Gehler, Sergio and Park, Wan-Il and Pato, Miguel",
    title = "{Gamma-ray boxes from axion-mediated dark matter}",
    eprint = "1303.6632",
    archivePrefix = "arXiv",
    primaryClass = "hep-ph",
    doi = "10.1088/1475-7516/2013/05/016",
    journal = "JCAP",
    volume = "05",
    pages = "016",
    year = "2013",
    note = "[Erratum: JCAP 03, E01 (2016)]"
}

@article{Klangburam:2023vjv,
    author = "Klangburam, Tanech and Pongkitivanichkul, Chakrit",
    title = "{Bounds on ALP-mediated dark matter models from celestial objects}",
    eprint = "2311.15681",
    archivePrefix = "arXiv",
    primaryClass = "hep-ph",
    doi = "10.1007/JHEP10(2024)145",
    journal = "JHEP",
    volume = "10",
    pages = "145",
    year = "2024"
}

@article{Yang:2024jtp,
    author = "Yang, Meiwen and Guo, Zhi-Qi and Luo, Xiao-Yi and Shen, Zhao-Qiang and Xia, Zi-Qing and Lu, Chih-Ting and Tsai, Yue-Lin Sming and Fan, Yi-Zhong",
    title = "{Searching accretion-enhanced dark matter annihilation signals in the Galactic Centre}",
    eprint = "2407.06815",
    archivePrefix = "arXiv",
    primaryClass = "hep-ph",
    doi = "10.1007/JHEP10(2024)094",
    journal = "JHEP",
    volume = "10",
    pages = "094",
    year = "2024"
}

@article{Calibbi:2025rpx,
    author = "Calibbi, Lorenzo and Li, Tong and Mukherjee, Lopamudra and Schmidt, Michael A.",
    title = "{Is Dark Matter the origin of the $B\to K \nu\bar\nu$ excess at Belle\textasciitilde{}II?}",
    eprint = "2502.04900",
    archivePrefix = "arXiv",
    primaryClass = "hep-ph",
    month = "2",
    year = "2025"
}

@article{Lee:2025jky,
    author = "Lee, Jong-Phil",
    title = "{New physics effects in $R(K^{(*)})$, $B_s\to\mu^+\mu^-$, and $B^+\to K^+\nu{\bar\nu}$}",
    eprint = "2502.06370",
    archivePrefix = "arXiv",
    primaryClass = "hep-ph",
    month = "2",
    year = "2025"
}

@article{Acanfora:2023gzr,
    author = "Acanfora, Francesca and Franceschini, Roberto and Mastroddi, Alessio and Redigolo, Diego",
    title = "{Fusing photons into nothing, a new search for invisible ALPs and Dark Matter at Belle II}",
    eprint = "2307.06369",
    archivePrefix = "arXiv",
    primaryClass = "hep-ph",
    doi = "10.1007/JHEP11(2024)156",
    journal = "JHEP",
    volume = "11",
    pages = "156",
    year = "2024"
}

@article{BaBar:2021ich,
    author = "Lees, J. P. and others",
    collaboration = "BaBar",
    title = "{Search for an Axionlike Particle in $B$ Meson Decays}",
    eprint = "2111.01800",
    archivePrefix = "arXiv",
    primaryClass = "hep-ex",
    reportNumber = "BABAR-PUB-21/006, SLAC-PUB-17631",
    doi = "10.1103/PhysRevLett.128.131802",
    journal = "Phys. Rev. Lett.",
    volume = "128",
    number = "13",
    pages = "131802",
    year = "2022"
}

@article{Wang:2021uyb,
    author = "Wang, Daohan and Wu, Lei and Yang, Jin Min and Zhang, Mengchao",
    title = "{Photon-jet events as a probe of axionlike particles at the LHC}",
    eprint = "2102.01532",
    archivePrefix = "arXiv",
    primaryClass = "hep-ph",
    doi = "10.1103/PhysRevD.104.095016",
    journal = "Phys. Rev. D",
    volume = "104",
    number = "9",
    pages = "095016",
    year = "2021"
}

@article{He:2025jfc,
    author = "He, Xiao-Gang and Ma, Xiao-Dong and Tandean, Jusak and Valencia, German",
    title = "{$B\to K\sf{+}invisible$, dark matter, and $CP$ violation in hyperon decays}",
    eprint = "2502.09603",
    archivePrefix = "arXiv",
    primaryClass = "hep-ph",
    month = "2",
    year = "2025"
}

@article{Berezhnoy:2025tiw,
    author = "Berezhnoy, Alexander and Lucha, Wolfgang and Melikhov, Dmitri",
    title = "{Analysis of $q_\mathrm{rec}^2$-distribution for $B\to K M_X$ and $B\to K^* M_X$ decays in a scalar-mediator dark-matter scenario}",
    eprint = "2502.14313",
    archivePrefix = "arXiv",
    primaryClass = "hep-ph",
    month = "2",
    year = "2025"
}

@article{Gubernari:2023puw,
    author = "Gubernari, Nico and Reboud, M{\'e}ril and van Dyk, Danny and Virto, Javier",
    title = "{Dispersive analysis of $B \to K^{(*)}$ and $B_{s} \to \phi$ form factors}",
    eprint = "2305.06301",
    archivePrefix = "arXiv",
    primaryClass = "hep-ph",
    reportNumber = "EOS-2023-02, IPPP/23/22, P3H-23-026, SI-HEP-2023-09",
    doi = "10.1007/JHEP12(2023)153",
    journal = "JHEP",
    volume = "12",
    pages = "153",
    year = "2023",
    note = "[Erratum: JHEP 01, 125 (2025)]"
}

@article{NA62:2020pwi,
    author = "Cortina Gil, Eduardo and others",
    collaboration = "NA62",
    title = "{Search for $\pi^0$ decays to invisible particles}",
    eprint = "2010.07644",
    archivePrefix = "arXiv",
    primaryClass = "hep-ex",
    reportNumber = "CERN-EP-2020-193",
    doi = "10.1007/JHEP02(2021)201",
    journal = "JHEP",
    volume = "02",
    pages = "201",
    year = "2021"
}

@article{BaBar:2013npw,
    author = "Lees, J. P. and others",
    collaboration = "BaBar",
    title = "{Search for $B \to K^{(*)} \nu \overline \nu$ and invisible quarkonium decays}",
    eprint = "1303.7465",
    archivePrefix = "arXiv",
    primaryClass = "hep-ex",
    reportNumber = "BABAR-CONF-13-002, SLAC-PUB-15411, BABAR-PUB-13-002",
    doi = "10.1103/PhysRevD.87.112005",
    journal = "Phys. Rev. D",
    volume = "87",
    number = "11",
    pages = "112005",
    year = "2013"
}

@article{Wang:2025tdx,
    author = "Wang, Yu-Ning and Duan, Xin-Chen and Tang, Tian-Peng and Wang, Ziwei and Tsai, Yue-Lin Sming",
    title = "{Exploring sub-GeV dark matter via $s$-wave, $p$-wave, and resonance annihilation with CMB data}",
    eprint = "2502.18263",
    archivePrefix = "arXiv",
    primaryClass = "hep-ph",
    month = "2",
    year = "2025"
}

@article{Belanger:2025kce,
    author = "B\'elanger, Genevi\`eve and Chakraborti, Sreemanti and Delaunay, C\'edric and Jomain, Margaux",
    title = "{Rekindling s-Wave Dark Matter Annihilation Below 10GeV with Breit-Wigner Effects}",
    eprint = "2503.08897",
    archivePrefix = "arXiv",
    primaryClass = "hep-ph",
    reportNumber = "IPPP/25/14, LAPTH-009/25",
    month = "3",
    year = "2025"
}

@article{Parrott:2022zte,
    author = "Parrott, W. G. and Bouchard, C. and Davies, C. T. H.",
    collaboration = "HPQCD",
    title = "{Standard Model predictions for B\textrightarrow{}K\ensuremath{\ell}+\ensuremath{\ell}-, B\textrightarrow{}K\ensuremath{\ell}1-\ensuremath{\ell}2+ and B\textrightarrow{}K\ensuremath{\nu}\ensuremath{\nu}\textasciimacron{} using form factors from Nf=2+1+1 lattice QCD}",
    eprint = "2207.13371",
    archivePrefix = "arXiv",
    primaryClass = "hep-ph",
    doi = "10.1103/PhysRevD.107.014511",
    journal = "Phys. Rev. D",
    volume = "107",
    number = "1",
    pages = "014511",
    year = "2023",
    note = "[Erratum: Phys.Rev.D 107, 119903 (2023)]"
}

@article{Kolay:2024wns,
    author = "Kolay, Lipika and Nandi, Soumitra",
    title = "{Exploring constraints on Simplified Dark Matter model through flavour and electroweak observables}",
    eprint = "2403.20303",
    archivePrefix = "arXiv",
    primaryClass = "hep-ph",
    doi = "10.1007/JHEP10(2024)008",
    journal = "JHEP",
    volume = "10",
    pages = "008",
    year = "2024"
}

@article{Kolay:2025jip,
    author = "Kolay, Lipika and Nandi, Soumitra",
    title = "{Flavour and Electroweak Precision Constraints on a Simplified Dark Matter Model with a Light Spin-0 Mediator}",
    eprint = "2503.15609",
    archivePrefix = "arXiv",
    primaryClass = "hep-ph",
    month = "3",
    year = "2025"
}

@article{Wang:2023trd,
    author = {Wang, Zeren Simon and Dreiner, Herbert K. and G{\"u}nther, Julian Y.},
    title = "{The decay $B\rightarrow K+\nu +\bar{\nu }$ at Belle II and a massless bino in R-parity-violating supersymmetry}",
    eprint = "2309.03727",
    archivePrefix = "arXiv",
    primaryClass = "hep-ph",
    doi = "10.1140/epjc/s10052-025-13745-6",
    journal = "Eur. Phys. J. C",
    volume = "85",
    number = "1",
    pages = "66",
    year = "2025"
}

@article{Bauer:2021mvw,
    author = "Bauer, Martin and Neubert, Matthias and Renner, Sophie and Schnubel, Marvin and Thamm, Andrea",
    title = "{Flavor probes of axion-like particles}",
    eprint = "2110.10698",
    archivePrefix = "arXiv",
    primaryClass = "hep-ph",
    reportNumber = "MITP/21-025, CERN-TH-2021-148, IPPP/21/37",
    doi = "10.1007/JHEP09(2022)056",
    journal = "JHEP",
    volume = "09",
    pages = "056",
    year = "2022"
}

\end{document}